\documentclass[modern, tighten, longauthor]{aastex63} 
\usepackage{xcolor, fontawesome, microtype}

\definecolor{twitterblue}{RGB}{64,153,255}
\definecolor{linkcolor}{rgb}{0.1216,0.4667,0.7059}
\def\wotan{{\tt W\={o}tan}}

\newcommand{\twitter}[1]{\href{https://twitter.com/#1}{\textcolor{twitterblue}{\faTwitter}\,\tt \textcolor{twitterblue}{@#1}}}

\shorttitle{\wotan: Comprehensive time-series de-trending in Python}
\shortauthors{Hippke et al.}

\begin{document}
\title{\wotan:\\Comprehensive time-series de-trending in Python with applications to exoplanet transit surveys}
\author[0000-0002-0794-6339]{Michael Hippke}
\affiliation{Sonneberg Observatory, Sternwartestr. 32, 96515 Sonneberg, Germany \twitter{hippke}}
\email{michael@hippke.org}

\author[0000-0001-6534-6246]{Trevor J. David}
\affiliation{Jet Propulsion Laboratory, California Institute of Technology,\\4800 Oak Grove Drive, Pasadena, CA 91109, USA}

\author[0000-0002-1078-9493]{Gijs D. Mulders}
\affiliation{Department of the Geophysical Sciences, University of Chicago \twitter{GijsMulders}}
\affiliation{Earths in Other Solar Systems Team, NASA Nexus for Exoplanet System Science}

\author[0000-0002-9831-0984]{Ren\'{e} Heller}
\affiliation{Max Planck Institute for Solar System Research,\\Justus-von-Liebig-Weg 3, 37077 G\"ottingen, Germany \twitter{DrReneHeller}}

\begin{abstract}
The detection of transiting exoplanets in time-series photometry requires the removal or modeling of instrumental and stellar noise. While instrumental systematics can be reduced using methods such as pixel level decorrelation, removing stellar trends while preserving transit signals proves challenging. Due to vast archives of light curves from recent transit surveys, there is a strong need for accurate automatic detrending, without human intervention. A large variety of detrending algorithms are in active use, but their comparative performance for transit discovery is unexplored. We benchmark all commonly used detrending methods against hundreds of \textit{Kepler}, \textit{K2}, and \textit{TESS} planets, selected to represent the most difficult cases for systems with small planet-to-star radius ratios. The full parameter range is explored for each method to determine the best choices for planet discovery. We conclude that the ideal method is a time-windowed slider with an iterative robust location estimator based on Tukey's biweight. This method recovers 99\,\% and 94\,\% of the shallowest \textit{Kepler} and \textit{K2} planets, respectively. We include an additional analysis for young stars with extreme variability and conclude they are best treated using a spline-based method with a robust Huber estimator. All stellar detrending methods explored are available for public use in \wotan, an open-source Python package on GitHub ({\color{linkcolor}\faGithub \,\url{https://github.com/hippke/wotan}}).
\end{abstract}

\NewPageAfterKeywords

\section{Introduction}
\label{sub:intro}
During the last decade, millions of stellar light curves have been collected with ground-based surveys (e.g.
HATNet, \citealt{2004PASP..116..266B};
WASP, \citealt{2006PASP..118.1407P};
KELT, \citealt{2007PASP..119..923P};
CHESPA, \citealt{2018arXiv180901789Z}) and from space (e.g., with
CoRoT, \citealt{2009A&A...506..411A};
\textit{Kepler}, \citealt{2010Sci...327..977B};
\textit{K2}, \citealt{2014PASP..126..398H})
and more missions are underway (such as
\textit{TESS}, \citealt{2015JATIS...1a4003R};
PLATO, \citealt{2014ExA....38..249R}).
These data are searched for tiny dips in brightness, potentially indicative of transiting planets. Due to the large number of light curves, search algorithms must be automatic, ideally performing detrending simultaneously with the transit search to avoid detrending-induced distortions. Alternatively, the most common method to find transits in stellar light curves with variability is to ``pre-whiten'' the data \citep{2004MNRAS.350..331A}, an approach that is supposed to remove all time-dependent noise (or irrelevant signals) from the data prior to a transit search. That said, no detrending algorithm is perfect and in fact they can actually induce transit-like false positive signals in the light curves that were not genuinely present in the data \citep{2018A&A...617A..49R}. We are left with the question as to which detrending algorithm works best for a subsequent transit search.

Instrumental trends in these data are usually mitigated by
fitting and
subtracting cotrending basis vectors \citep{2016ksci.rept....9T},
decorrelation techniques between photometrically extracted light curves and the telescope pointing \citep{2014PASP..126..948V,2015ApJ...806...30L,2016MNRAS.459.2408A},
or methods such as the trend filtering algorithm \citep[TFA,][]{2005MNRAS.356..557K} using simultaneously observed stars to measure and remove systematic effects, and similar procedures \citep{2009MNRAS.397..558K}.

\begin{figure*}
\includegraphics[width=\linewidth]{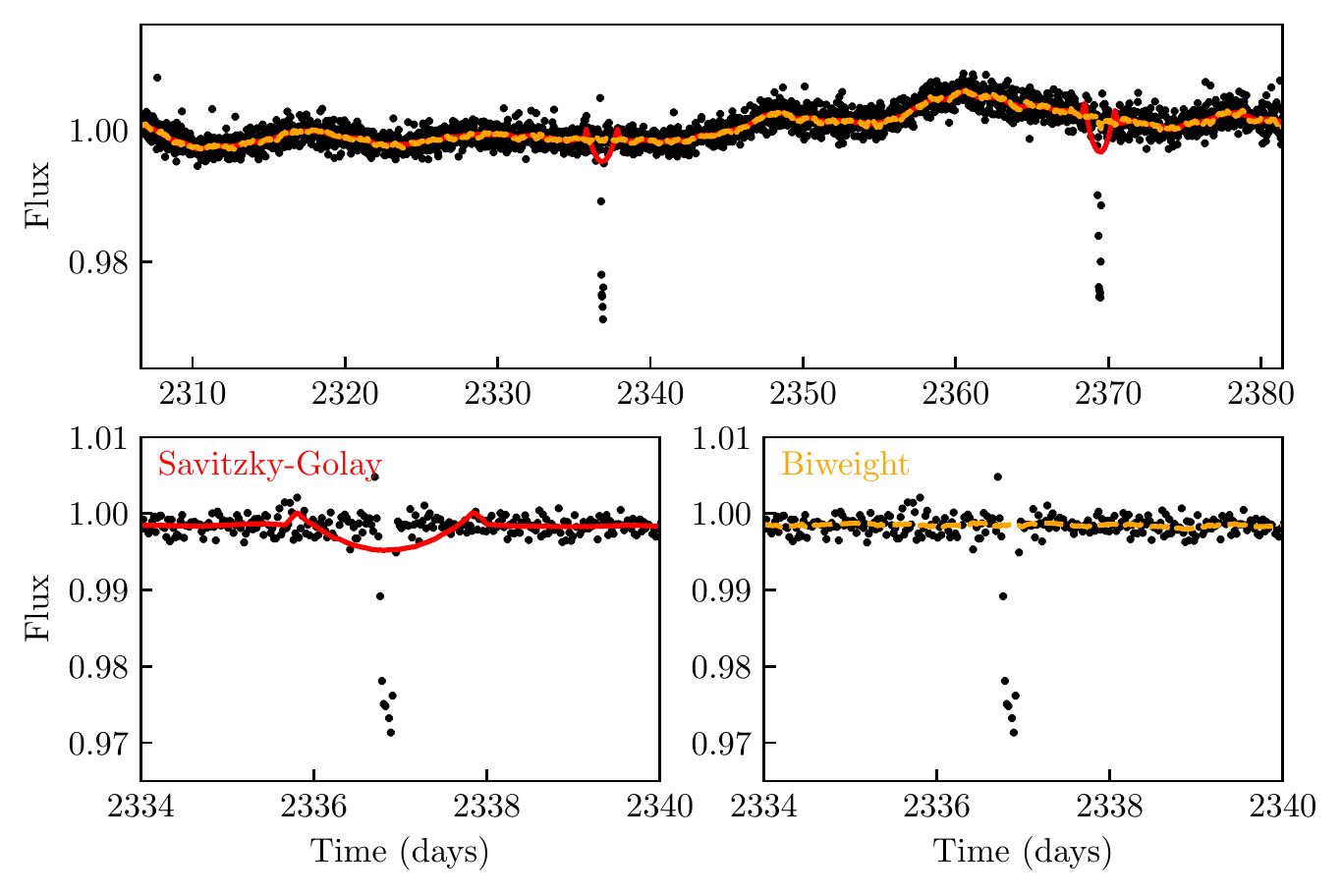}
\caption{Light curve of EPIC211995398 \citep[black points,][]{2016MNRAS.461.3399P} with a Savitzky-Golay filter (red) as used in the {\tt lightkurve} package \citep{https://doi.org/10.5281/zenodo.1181928} with a default of $w=101$ cadences, polynomial order $p=2$, and a time-windowed estimator (orange, Tukey's biweight, $w=0.5\,$d). The latter represents a reasonable fit to the data, while the Savitzky-Golay overfits the in-transit points and reduces the transit depth by $\sim15\,$\%, making detection more difficult (although in this example still trivial).}
\label{fig:fail}
\end{figure*}

However, even post-instrumental processing, transit dips are often overshadowed by stellar trends \citep{2015ApJ...810...29H}. Stellar noise is star-specific and occurs with diverse characteristics on all time scales. Stellar activity cycles, such as the 11-year solar activity cycle \citep{1844AN.....21..233S,2009ApJ...700L.154U}, appear to be present on different time scales in other stars as well \citep{2010Sci...329.1032G,2017ApJ...851..116M}. The solar rotation of $\sim27\,$\,d \citep{1934TeMAE..39..201B,2000SoPh..191...47B} produces noise from spots, as discovered by Galileo Galilei in 1612. Other stars exhibit brightness variations on many different time scales and with a range of amplitudes. Young stars can show bursts with amplitudes of up to 700\,\% on time scales of 1\,d to 10\,d \citep{2017ApJ...836...41C,2018AJ....156...71C}, and RRab Lyrae have amplitudes of $\sim$50\% over 0.5\,d. Photometrically quiet stars like our Sun or Kepler-197 \citep{2014ApJ...784...45R}, exhibit only weak (0.1\%) long-time (months) variation. Intermediate cases are common, e.g. Kepler-264\,b \citep{2015ApJ...806...51H} with trends on time scales of hours to days. Systems where stellar variability and transit signals occur on similar timescales require robust detrending methods, see Figure~\ref{fig:fail} as an example of filter-size dependency, where the transit depth is altered when using the Savitzky-Golay method.

The importance of variability on short time scales has motivated the usage of a broad range of different detrending methods. After an extensive literature research, however, it appears that there is no systematic overview of their strengths and weaknesses.\footnote{That said, we found a brief comparison of a sliding median with a sum-of-200-sines filter by \citet{2009A&A...495..647B}. And \citet{2018A&A...617A..49R} investigated the suitability of various detrending methods in the context of exomoon detections.} Common methods are sliding medians \citep[e.g.,][]{2009Natur.462..891C,2013A&A...553A..30T,2014ApJS..210...19B, 2016ApJ...827...78S,2019arXiv190109930D},
sliding means \citep{2010arXiv1004.0836W,2010ApJ...716..315M},
polynomial filters \citep{2012ApJ...750..114F,2012ApJ...749...15G,2018MNRAS.475.1809G},
splines \citep{2016ApJS..222...14V,2016ApJ...818...46M, 2018AJ....156...78L,2018AJ....155..136M,2019MNRAS.484.3731R},
LOESS regressions \citep{2017NatAs...1E.129L,2019arXiv190105116C},
Cosine Filtering with Autocorrelation Minimization \citep[CoFiAM,][]{2013ApJ...770..101K,2018A&A...617A..49R},
Gaussian processes \citep{2015MNRAS.447.2880A,2016ApJS..226....7C,2019ApJ...870L..17C},
the \citet{1964AnaCh..36.1627S} filter \citep{2011ApJ...729...27B,2011ApJS..197....6G,2012PASP..124..985S,2014PASP..126..398H},
wavelets \citet{2009ApJ...704...51C,2013Sci...342..331H,2013ApJS..204...24B,2015ApJ...800...46B,2015ApJS..217...18S,2016arXiv160708417G},
frequency filtering through Fourier decomposition \citep{2013ApJ...767..137Q},
sometimes as a Butterworth filter \citep{2018arXiv181209227N},
and combinations such a sliding median followed by a polynomial, \citep{2016ApJ...829...23D,2018ApJ...858...55P}.
and sliding median filters \citep{2019A&A...625A..31H}.
Occasionally, adaptive filtering is performed. After determining the main modulation period from stellar rotation, a specific window (or kernel) can be defined \citep{2016MNRAS.461.3399P,2017AJ....154..224R}.

This myriad of methods opens the valid question of what works best for the purpose of blind transit searches, and how big the discrepancy is in using the best, versus some other method. Here, we present a review and comparison of the relevant methods for removing stellar variability. The algorithms presented have been included in \wotan, an open-source Python package. 

An alternative approach to detrending with a subsequent transit search is the \textit{simultaneous} modeling of trends with a transit search. This approach was applied by \citet{2015ApJ...806..215F} to one K2 campaign, suggesting that the computational cost is acceptable. Recent speed-ups from linear algebra and gradient-based methods introduced by new tools such as Starry \citep{2018ascl.soft10005L,2019AJ....157...64L} and Celerite \citep{2017AJ....154..220F} may make this approach more attractive. It is, however, still unclear whether the simultaneous method is beneficial. A recent analysis concluded that  ``the simple (and more traditional) method that performs correction for systematics first and signal search thereafter, produces higher signal recovery rates on the average`` \citep{2016A&A...585A..57K}.

The paper is structured as follows: In Section~\ref{sec:algo}, we review the commonly used estimators and detrending methods. In Section~\ref{sec:tests}, we present three experiments to test these methods, with the results in Section~\ref{sec:results}.

\section{Algorithms}
\label{sec:algo}
Common algorithms (Section~\ref{sub:intro}) can broadly be categorized as sliding filters, splines and polynomials, and Gaussian processes. This section reviews the relevant methods.

\subsection{Sliding filters}
\label{sec:sliding}

In the statistical literature, a common filter like a sliding (rolling, walking, running) mean or median is categorized as a {\it scatterplot smoother} in the form

\begin{equation}
y = s(x) + \epsilon
\end{equation}

\noindent
where a trend (or smooth) $s$ is estimated. After sorting the data in chronological order so that $x_1<x_2< \dots < x_n$, a simple (cadence based) sliding mean smoother $S$ can be calculated for each $x_i$ by averaging a range of $y_j$ values corresponding to each $x_i$:

\begin{equation}
S(x_i)= \sum_{j \in W(x_i)} (y_j) / w_i
\end{equation}

\noindent
in a neighborhood with indices $W(x_i)$ that contains a total of $w_i$ data points. Typically, the neighborhood is symmetrical with the nearest $2k+1$ points, including $k$ data points taken before $x_i$, $k$ data points taken after $x_i$, and $x_i$ itself:

\begin{eqnarray}
W(x_i) = \{ {\rm max} (i-k,1) \dots , i-1, i, i+1, \dots, {\rm min}(i+k,n) \} \ .
\end{eqnarray}

This kernel smoother uses a rectangular window, or a ``box'' (sometimes called tophat, or boxcar) kernel. Other commonly used kernels are Gaussian densities (computationally expensive because they never reach zero) and e.g., the \citet{Epanechnikov1969}, a parabola. To calculate $W(x_i)$ with a non-rectangular kernel, the $j$-th point is given a weight

\begin{eqnarray}
g_{ij} = \frac{c_i}{\lambda} d \left( \frac{|x_i - x_j|}{\lambda} \right) \ ,
\end{eqnarray}

\noindent
where $\lambda$ is a bandwidth tuning constant, $c_i$ is used to normalize the sum of the weights to unity for each $x_i$, and $d$ describes the kernel. For example, the Epanechnikov kernel has $d(z)=3/4 (1-z^2)$ for $z^2<1$, (with $z$ as the distance from the midpoint of the kernel) and zero otherwise.

Cadence-based filters are applied to a fixed number of data points, whereas time-windowed filters use a window fixed in a second dimension. Cadence-based filters are simple and fast (Section~\ref{speed}), but require equally spaced data without gaps to function as designed. In the real world, data points are often missing due to e.g., cosmic ray hits or momentum dumps. Their spacing in time typically originates with a constant cadence of a {\it local} clock. On a body moving around the sun in an Earth-like orbit, barycentering these observations causes cadence spacings to periodically drift by 16 minutes over the course of a year. The most severe effect for cadence-based filters in practice, however, are longer gaps, e.g. a cadence-based filter stitching together sinusoidal variation where half a rotation period is missing.

\subsection{Cadence-based sliding filters}
For cadence-based sliders, we test a sliding median \citep[using the {\tt scipy.medfilt} implementation,][]{scipy:2001} and the \citet{1964AnaCh..36.1627S} method {\tt scipy.savgol\_filter}. It fits successive sub-sets of equally-spaced points with a low-degree polynomial (typically order $p=2$ or $p=3$) with a least-squares calculation. The resulting analytical solution is then applied to all subsets. The Savitzky-Golay is the default detrending method in the {\tt lightkurve} package \citep{https://doi.org/10.5281/zenodo.1181928} with a default of $w=101$ cadences and $p=2$. Following its original definition, it is cadence-based, and we are not aware of any other usage.

\subsection{Time-windowed sliding filters}
A time-windowed filter avoids issues faced by cadence-based ones. We can define a time-windowed filter with a window length $w$ centered on the time $t(x_i)$ for each $x_i$. A rectangular (i.e., symmetric and with uniform weights) window of size $w$ contains those data points $N(x_j)$ whose $t(x_j)$ are in $[t(x_j)-w/2 \dots t(x_j)+w/2 ]$. We implement a module for such a time-based slider in {\tt Python} to allow for a comparison of various estimators from the astrophysical and statistical literature.

\begin{figure*}
\centering
\includegraphics[width=\linewidth]{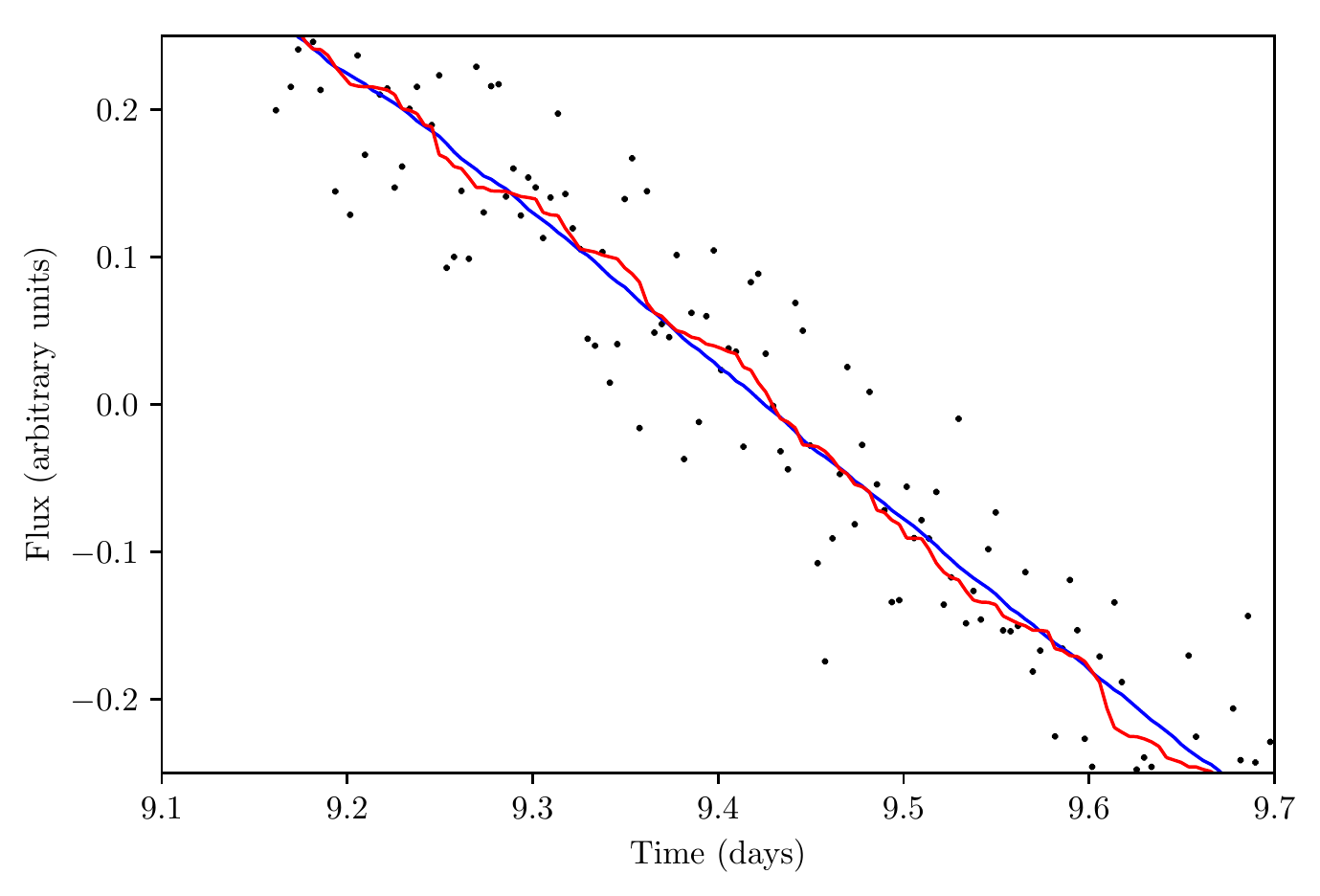}
\caption{Synthetic white noise data (black) with a sliding mean (blue) and sliding median (red) trend, both of the same window size of 101 cadences. The sliding mean is very close to the true trend, while the median shows additional jitter, because its estimate is 25\,\% less efficient.}
\label{fig:mean_median}
\end{figure*}

\subsubsection{Mean and median}
The most common smooth location estimators at each time $t(x_j)$ are the mean and the median. These two represent the extreme ends of estimators with respect to robustness and efficiency. For a normal distribution, the mean is the most efficient estimator. A maximally efficient estimator is defined as the one which has the minimal possible variance of the location estimate (the minimum mean squared error in the unbiased case) for a given noise distribution \citep[][\S\S\,3.2, 4.8f]{kenney1947mathematics}. The median is robust for distributions with up to 50\,\% outliers, but its efficiency, measured as the ratio of the variance of the mean to the variance of the median, is lower. The uncertainty on a sample median is $\sqrt{\pi /2}\approx 1.25$ times larger than the uncertainty on a sample mean \citep[][p.~694]{Huang1999,press1992numerical}. The mean, however, can be pushed towards $\pm \infty$ with a single outlier. This is known as the {\it bias-variance tradeoff}. The median results in maximal robustness, but minimal efficiency -- this extra source of jitter can be seen by-eye (Figure~\ref{fig:mean_median}).
Another issue with the median is visual: Its trend is always {\it exactly} equal to one of the points inside the sliding window. Dividing the flux by the trend, and plotting these detrended data, results in a large number of points equal to unity. This cluster of points parallel to the abscissa in a diagram may appear visually unprofessional and disqualify the median for publication quality Figures.

There is a rich statistical literature about robust estimators in between these two extremes, each maximizing some ratio of efficiency versus robustness. For example, the trimmed mean is a simple robust estimator of location that deletes a certain percentage of observations (e.g., 10\,\%) from each end of the data, then computes the mean in the usual way. It often performs well compared to the median and mean, but better robust estimates are available for many use cases. The following subsections review some of these estimators.

\subsubsection{Trimmed mean and median}
A very simple robust location estimator is the trimmed mean
which rejects a small fraction of the most extreme values in a time series of observations prior to averaging the remaining data. This brings robustness over the simple mean, while maintaining most data, allowing for a better efficiency compared to the median. When this clipping is increased, it becomes intuitively clear that the median is the result of clipping 50\,\% of each side of the distribution. A one-sided x-sigma clipper is expected to remove

\begin{equation}
Y=\frac{2}{1-{\rm erf} \left( \frac{x}{\sqrt{2}}\right)}
\end{equation}

\noindent
values from a Gaussian distribution, where ${\rm erf}(x)$ is the Gauss error function. For $x=3\,\sigma$, where $\sigma$ is the standard deviation of the sample, we expect to clip $\sim1/741$ values, or $\sim5$ from a typical \textit{K2} light curve worth $3{,}500$ data points. In practice, often many more outliers are present. One issue with the truncated mean is that it produces a biased estimator if the underlying sample is not symmetric. In the presence of transits (and few flares), this is often the case.

In effect, a trimmed mean is a two-step procedure. First, outliers are detected and removed. Second, the efficient estimation method (the least squares) is applied to the remaining data. There are two general challenges with this approach. First, the outlier detection method relies on a non-robust initial estimate. This can result in the masking effect, where a group of outliers masks each others and escapes detection \citep{Rousseeuw1987}. Second, if this is avoided with a high breakdown initial fit (i.e., a larger proportion of the data is removed), the following estimate is inefficient (similar to the median), plus it inherits the inefficiency of the initial (non-robust) estimate \citep{He1992}.

\subsection{The Huber function}
While the trimmed mean performs well, compared to the median and mean, better robust estimators are often available. Peter J. Huber proposed a generalization of the maximum likelihood estimation called M-estimators \citep{Huber1964}. Non-linear least squares and maximum likelihood estimation are special cases of M-estimators. Given an assumed distribution, one can select an estimator with the desired properties for the bias-variance tradeoff. The estimator is optimal {\it if the data are close to the assumed distribution}.

Estimators can be distinguished by their {\it loss function}. The ordinary least squares method uses the squared loss ($L(a)=a^2$), where $a$ are the residuals, i.e. the difference between the observed values and the midpoint. A robust estimator with a loss function that is less affected by very large residual values has been described by \citet{huber1981robust}, \citet[][p.~172]{huber2011robust}. The Huber function is quadratic for small values, and becomes linear for larger values. Its loss function can be approximated as \citep{Charbonnier1997}

\begin{equation}
L (a) = c^2\left(\sqrt{1+(a/c)^2}-1\right) \ ,
\end{equation}

\noindent
where $c$ is a tuning parameter to adjust the steepness of the transition between quadratic loss (mean squared error, $a^2/2$ for small values) and linear loss (absolute deviation, $a$ for large values). For $c \rightarrow \infty$, the function becomes the usual least-squares estimator.

\begin{figure}
\centering
\includegraphics[width=.49\linewidth]{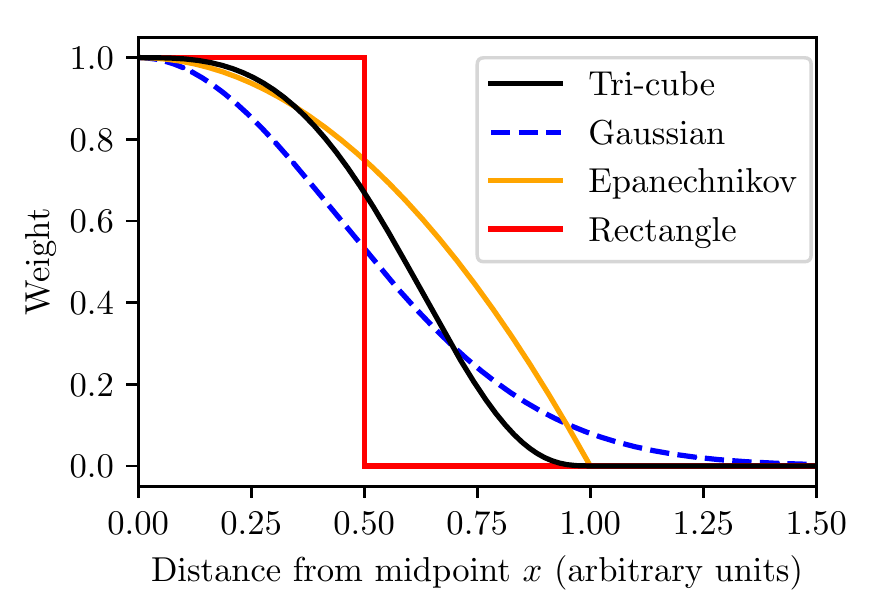}
\includegraphics[width=.49\linewidth]{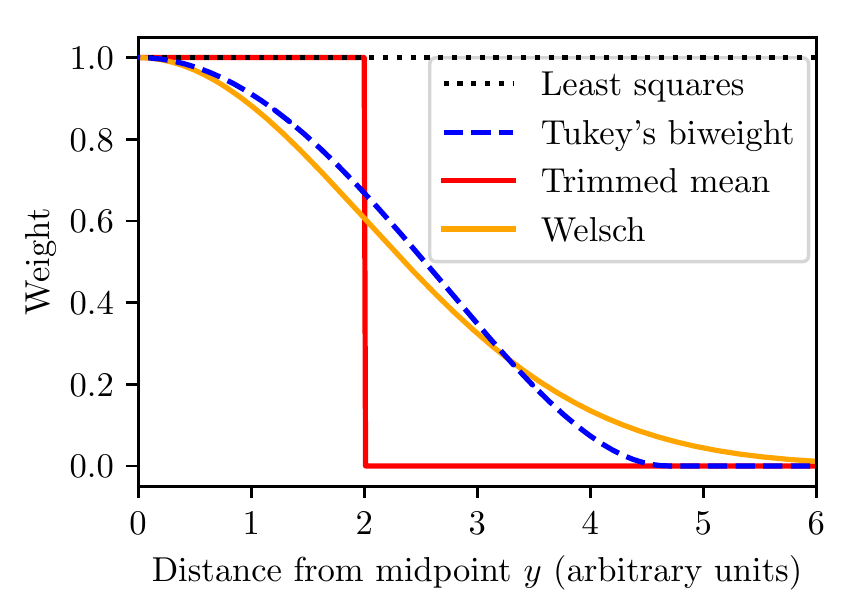}
\caption{Left: Common window functions for weights of points of the independent variable (usually $x$, or $t$ in our case). Most commonly, a rectangular window (red line) assigns weights of unity to all values within the current (sliding) window, and zero otherwise. Other well-known choices are the Gaussian and the Epanechnikov. The tri-cube is somewhat steeper in slope, giving more weight to the center (but not as much as the rectangle). Right: Relative weights of robust estimators as a function of distance from the mid-point $y$ (here: flux), where the ordinary least-squares method has constant weight. The trimmed mean is a step-function shown for a cut fraction of 0.1, which corresponds to a $\sim2\,\sigma$ clipping in a Gaussian distribution. The estimators by Tukey (here $c=4.685$) and Welsch ($c=2.11$) have smoother transitions.}
\label{fig:weights}
\end{figure}

\subsubsection{Tukey's biweight (bisquare)}
A well-known robust location estimator is Tukey' biweight (or bisquare) \citep[][pp.~203-209]{mosteller1977data}\footnote{This method is named after the mathematician John Tukey (1915--2000) who also coined the terms ``bit`` \citep{Shannon1948} and ``software`` \citep{Buchholz2000}, invented the Fast Fourier Transform \citep{Cooley1965}, and the boxplot \citep{tukey1977exploratory,McGill1978}, amongst many other things.}. Compared to the ordinary least-squares, where the weights are constant at unity, the biweight has weights as a distance from the central location as $L(a) = (-(a/c)^2)^2$ for $|a|<c$, and zero otherwise (Figure~\ref{fig:weights}). Typical values are $c=4.685$ for an asymptotic efficiency of 95\,\% compared to that of least-squares, for a normal distribution. Larger values for $c$ make the estimate {\it less robust}, but {\it more efficient}. Another commonly found value is $c=6$ which includes data up to four standard deviations from the central location \citep{mosteller1977data}. Then, the efficiency is $\sim 98\,\%$ \citep{Kafadar1983}. The biweight with $c=6$ has proven useful for velocity clustering of galaxies \citep{1990AJ....100...32B}. Our tests with \textit{K2} data show $c\sim5$ to be optimal, although the variation for $c=4$ and $c=6$ result only in the loss of one planet (of 100) in the experiment in Section~\ref{sub:k2}.

The biweight can be used as a one-step estimate \citep[as implemented by {\tt Astropy};][]{2018AJ....156..123A} with the median as the first guess for the central location. Alternatively, it can be approximated with iterative algorithms such as Newton-Raphson, where the result of each iteration is used as the starting guess for the next. We have determined the difference between the one-step estimate and a converged solution numerically. This difference is typically small, but not negligible \citep[in agreement with][p. 152]{hampel2011robust}: usually between a few and a few tens of ppm ($10^{-6}$) in \textit{Kepler} and \textit{K2} data. This corresponds to a few percent of the depth of an Earth-Sun transit ($\sim84\,$ppm). This additional detrending noise, caused by a sub-optimal location estimator, is avoidable with a few (3--5) Newton-Raphson iterations for convergence to the $<10^{-6}$ ($<1\,$ppm) level. Our implementation within \wotan \, allows to set this threshold to an arbitrary value.

\subsubsection{Other robust estimates}
There is a variety of other robust M-estimators which share the idea that data near the center is assigned more weight. Among the most common are the Hodges-Lehmann-Sen \citep{Hodges1963,Sen1963} with a breakdown point (fraction of outliers above which results become incorrect) of 0.29 (median: 0.5). It is most suited if the underlying distribution is a mixture of normal distributions. Other estimators include the Welsch-Leclerc with an exponential decay, $L(a) = \exp((a/c)^2)$  \citep{Dennis1978,DBLP:journals/ijcv/Leclerc89}, and Andrew's sine wave \citep[][pp. 3-28]{10.2307/j.ctt13x12sw.3}. A related class of estimators is rank-based (using adaptive weights), aiming to maximize both robustness and efficiency. From this class, we include the $\tau$ estimator in the experiment \citep{doi:10.1080/01621459.1988.10478611}.

\subsection{Local regressions: LOWESS}
LOWESS/LOESS (locally weighted/estimated scatterplot smoothing) is a non-parametric regression technique developed by \citet{Cleveland1979,Cleveland1981}. In essence, the method is a generalization of a moving window, where each window is used to fit a locally-weighted polynomial regression (see Chapter 8 of \citet{chambers1992statistical}; \citet{Wilcox2017}). The window can be a boxcar in the $x$-axis, or any other window function (e.g., a Gaussian). The regression on $y$ is typically performed with least-squares or a robust estimator such as the biweight. For our tests, we use the implementation from the Python package {\tt statsmodels} \citep{seabold2010statsmodels} based on \citet{hastie2009elements} with a bisquare hardcoded to $c=6$ on the $y$ values, selectable iterative depth, and a tricube window in $x$ with weights as $(1-d^3)^3$ with distance $d$ from $x$. Similar parameters have been used to determine galaxy properties \citep{2013MNRAS.432.1862C}. \citet{2018MNRAS.475.1809G} re-invented a filter similar to the LOWESS with outlier-clipping for robustness, instead of a smooth biweight. LOWESS appears to be rarely used, with ony two ocurrences in a literature review \citep{2017NatAs...1E.129L,2019arXiv190105116C}.

\subsection{Friedman's ``Super-Smoother``}
\citet{friedman1984variable} proposed a regression smoother based on local linear regression with adaptive bandwidths. In multiple passes, it selects a ``best" bandwidth from initial estimates at each data point over the range of the predictor variable. For our tests, we use the implementation by \citet{2015zndo.....14475V,2015zndo.....28518V}. Finding appropriate duration ranges for the three rounds of iterations, and the ``bass enhancement" $\alpha$ which sets the ``smoothness" proves computationally expensive. A major problem for the ``Super-Smoother" in light curve data is the fact that it assumes Gaussian errors, which is problematic due to the presence of outliers (e.g., transits). The method has been used with success, however, for finding periods in eclipsing binaries \citep{2011ApJ...731...17B}.

\subsection{Gaussian processes}
Gaussian processes (GPs) are popular in many sub-field within astrophysics, such as
spectroscopy \citep[e.g.,][]{2012MNRAS.419.2683G},
pulsar timing \citep{2014PhRvD..90j4012V},
asteroseismology \citep{2009MNRAS.395.2226B},
redshift prediction \citep{2009ApJ...706..623W}, and
supernova luminosity \citep{2013ApJ...766...84K}.

GPs have also found an application in the simultaneous modelling of noise and transits \citep[e.g.,][]{2017AJ....154..254G}. Here, we only benchmark GPs against other available methods for our scope of removing stellar trends while preserving transits. The reason we do not perform simultaneous transit searches and GP fits (or any detrending, for that matter) is simple: it is prohibitively computationally expensive.

Usually, GPs are used assuming Gaussian noise. GPs can be used as priors with other (more robust) likelihood functions, such as the Student's t-distributed noise model \citep{1997physics...1026N}.

\cite{2018MNRAS.474.2094A} provides an example of using GPs to model and stellar rotation. First, a Lomb-Scargle periodogram is calculated to determine the rotation. A periodic GP is then informed about the strongest periodic signal. This method was partially successful ($\gtrsim80\,\%$) in fitting the test planets.

We test GPs using {\tt george} \citep{2015ITPAM..38..252A,2017AJ....154..220F} and {\tt celerite} \citep{2017ascl.soft09008F,2018RNAAS...2a..31F} with a squared-exponential kernel \citep{2013EPSC....8..599D} as used by \citet{2015ApJ...800...46B,2016ApJS..226....7C,2018ApJS..235...38T}, the Matern 3/2 \citep{2019AJ....157...51W}, and the quasi-periodic \citep{2018arXiv181102156B} for our test sample of young stars.

We test our GPs in the one-pass mode. In addition, we test an iterative $2\sigma$ outlier clipping from the fitted trend in each iteration until convergence. The GP hyper-parameters of each kernel are determined by iterating over a grid of plausible values, choosing those which maximize the planet detection yield.

\subsection{Splines}
\citet{schoenberg1946contributions} introduced ``splines`` as a smooth, piecewise polynomial approximation. Today, the most commonly used type are univariate (cardinal) basis (B-) splines \citep{R1980} where all knots lie in a single dimension and are equally spaced \citep{Ferguson1964,ahlberg1967theory}. Usually, splines are fit by minimizing the sum of the squared residuals. As with a sliding mean, the estimate is affected by outliers. A common method to increase the robustness is an iteratively sigma-clipping re-weighting procedure, also with least-squares calculations. In each iteration, outliers beyond some threshold are clipped, and the procedure is repeated until convergence. We implement this method based on the {\tt SciPy} package in order to offer segmentations of the light curve which has similar methods to the package {\tt untrendy}{\footnote{\url{https://github.com/dfm/untrendy}}}.

An alternative approach is to use a linear loss \citep{lawson1961contribution,rice1968lawson}, or a robust M-estimator such as the Huber or the biweight. In our test we include a spline estimated with a Huber-regressor, provided by the {\tt scikit-learn} package. 

With these methods, the knot distance must be chosen manually. As with window sizes, there is a trade-off between under- and overfitting. There exist a range of heuristic methods to determine a sensible knot spacing \citep{friedman1989flexible,kooperberg1991study}. The most established method, however, are penalized (P-) splines \citep{eilers1996}. The extra degrees of freedom from additional knots is judged against the smaller residuals from an improved fit. The optimal weight of the penalty can be determined using cross-validation and Bayesian arguments. We implement this automatic spline-fitter using {\tt pyGAM} \citep{danielserven}, which conveniently offers the creation of Generalized Additive Models and their selection. We add iterative sigma-clipping for $2\,\sigma$ outliers from the fitted trend at each iteration until convergence.

\subsection{Cosine Filtering with Autocorrelation Minimization}
The CoFiAM algorithm \citep{2013ApJ...770..101K} was developed for the application of exomoon-hunting, where it is crucial to protect a time window longer than the transit duration from method-induced trends. It was specifically optimized to reconstruct the morphology of a previously identified transit, and the surrounding region, as faithfully as possible. To do this, it is required to know the location and duration of the transits, and mask them, to assure that the trend is unaffected by the lower in-transit points. Thus, CoFiAM is not per se suited for our blind search experiment.

While the code was described in-depth in \citet{2013ApJ...770..101K}, it was not released as open source. However, it was re-written for validation purposes by \citet{2018A&A...617A..49R} following the description in the original paper. This implementation is also available through our \wotan\,\,package.

CoFiAM uses a sum-of-(co)sines approach, where the optimal detrending is determined iteratively. In each iteration, one harmonic order of a cosine is added and evaluated in the least-squares sense, so that the autocorrelation of the residuals is minimal. Autocorrelation is the correlation of the signal with a delayed copy of itself as a function of delay. CoFiAM employs the Durbin-Watson statistic to quantify the level of an autocorrelation with a lag of 30 minutes in the default setting. Many models with sums of cosines are tested, of increasing orders until these are too large so that the chosen window size would be compromised\footnote{A similar method, based on polynomials, was used by \citet{2019ApJ...870L..17C}.}. Afterwards, the model with the lowest autocorrelation is selected. 

The method can be made outlier-resistant by applying a sliding filter before the actual process. In the original implementation, a 20-cadence window is used with the median as a center estimate, clipping all points more than 3 standard deviations away. In our \wotan\,\,package, a feature-rich outlier-clipping method is available, offering a time-windowed slider with the mean or median as the center estimate, and the standard deviation or the median absolute deviation as the $\sigma$ clipper.

\subsection{Iterative sum of sines and cosines}
Similar to CoFiAM described in the previous section, a sum of sines and cosines can be fit to the data \citep{2010A&A...521L..59M}. Again, the highest degree can be chosen so that a defined window in time is protected. Now, however, no decision is made based on autocorrelation measurements. Instead, the data are divided by the trend, and outliers (e.g., more than 2 standard deviations from the local mean) are temporarily removed). The procedure is repeated iteratively until convergence, i.e. until no further clipping occurs. Afterwards, this trend is applied to the complete data set\footnote{We thank Aviv Ofir for suggesting this method.}.

\subsection{Edge treatment}
\label{sub:edge}
Any detrending method is challenged near the edges of a time series, where data is missing on one side. There are various methods to handle such situations. On the one hand, the data can be padded to enlarge a full window $w$ (or kernel). Padding can be done by adding a constant value, by mirroring the values nearest to the edge, or by wrapping the beginning and the end of the time series. On the other hand, data near the edge can be discarded after detrending. Some algorithms use yet other methods, such as the Savitzky-Golay filter in {\tt scipy.savgol\_filter}, which fits a polynomial to the end segments.

Dropping data near the edges removes at least $2 \times 0.5w \sim 1\,$d out of 80\,d \textit{K2} data ($\sim1\,\%$), and more if additional gaps during the time series occur.

\section{Test samples}
\label{sec:tests}
To cover a maximally diverse sample of transits, we decided to split the experiment into three sample groups. In each, we search for known transits (real or injected) using the Transit Least Squares ({\tt TLS}) transit detection algorithm \citep{2019A&A...623A..39H} with stellar parameters on temperature, surface gravity, radius, and from the EPIC \citep{2011AJ....142..112B}, KIC \citep{2014ApJS..211....2H,2016ApJS..224....2H,2017ApJS..229...30M} and TIC catalogs \citep{,2018AJ....156..102S}, cross-matched with theoretical limb-darkening values from \citet{2018A&A...618A..20C,2018yCat..36180020C}. This procedure is similar to the one applied by the transit least-squares survey \citep{2019A&A...625A..31H,TLSSII}. We set up {\tt TLS} with a fine search grid of the parameter space using an oversampling factor of five for the transit period and a transit duration grid spacing of 5\,\%, with all else as the default in {\tt TLS} version 1.0.23. A detection was accepted if the published period matched the detected period as detected by {\tt TLS} within 1\,\%, and if it was the strongest period detected. Usually, a fixed-value threshold is applied below which detections are not accepted, in order to limit the number of false positives due to noise. For {\tt TLS} and white noise, an SDE (signal detection efficiency) of 9 yields a false alarm probability of 0.01\,\%. An SDE value of $x$ for any given period means that the statistical significance of this period is $x\,\sigma$ compared to the mean significance of all other periods. This assumes Gaussian (white) noise.

A similar measure of significance is the signal-to-noise ratio which compares the depth of the transit model with the out-of-transit noise:
\begin{equation}
{\rm SNR} = \sqrt{N_{T}}\frac{T_{dep}}{\sigma_{OT}}
\end{equation}
with $N_{T}$ as the number of transit observations, $T_{dep}$ the transit depth and $\sigma_{OT}$ the standard deviation of out-of-transit observations. For Gaussian noise, SDE and SNR are qualitatively identical.

In reality, noise is often time-correlated (red), so that spurious signals mimic real transits. Then, the false alarm probability is underestimated. A metric which takes correlated noise into account is the signal-to-pink-noise ratio derived by \citet{2006MNRAS.373..231P}. It weights the SNR with the amount of red noise in the vicinity of the transit.

For our experiments presented here, we employ the most commonly used metric SDE. We have not applied a threshold, as we are not interested in absolute, but relative performances. The lower 10\,percentile of SDE values gives a useful insight into potentially missed planets due to a user-chosen cut-off.

Most previous transit searches used the BLS (Box Least Squares) algorithm \citep{2002A&A...391..369K}, where the search filter is not a transit-shaped curve, but a box. This reduces the SDE (and thus the recovery rate for small planets) by $\sim5-10\,\%$. Other than that, we expect no difference in the relative performance of the detrending algorithms. For a comparative analysis of the recovery rates for small planets, see Section 3.1 in \citet{2019A&A...623A..39H}.

\subsection{Robustness sample of real planets in \textit{K2}}
In the first set, we use 100 light curves from the \textit{Kepler} \textit{K2} mission taken in 13 of the 19 \textit{K2} campaigns in different target fields along the ecliptic between 2014 and 2018. Each light curve covers a time span of about 75\,d. We use data from the {\tt EVEREST} reduction \citep{2016AJ....152..100L,2018AJ....156...99L} which removes most of the instrumental noise. These data are relatively recent and are still being searched for additional planets \citep{2019A&A...625A..31H,TLSSII}. While most systematics are removed by {\tt EVEREST}, some light curves are still contaminated with unflagged outliers. These (low flux) outliers, together with red-noise systematics, make detrending difficult.

To test a wide variety of parameters, we select candidates from the NASA Exoplanet Archive\footnote{\url{https://exoplanetarchive.ipac.caltech.edu/}} \citep{2013PASP..125..989A}. We include planets among the largest and smallest transit depths, and those with the longest and shortest transit durations (but requiring $n\geq3$ transits). The remainders were selected randomly for a total of 100 light curves of planet candidates.

The orbital periods ($P$) in this set range between 0.6 and 35.4\,d, with a mean of 10.4\,d and a median of 7.4\,d. The transit depths are between 0.009\,\% (96\,ppm) and 5\,\% with an mean of 0.5\,\% and a median of 0.1\,\%. The transit durations ($T_{14}$) from first to fourth contact \citep{2003ApJ...585.1038S} cover the range 0.004--0.14\,$T_{14}/P$ (mean 0.02, median 0.01).

\subsection{Performance sample of shallowest planets in \textit{Kepler}}
The original \textit{Kepler} mission provided more consistent data quality (compared to \textit{K2}) with overall lower noise and, more importantly, few low outliers or other artefacts. It is an effective test bed for a search of the shallowest transiting planets.

Among the planets most difficult to find are those with the shallowest transits. These are also of high interest, as they contain the smallest (Earth-sized), rocky planets. To test the optimal detrending for the shallowest planets, we select the 100 validated planets with the lowest signal-to-noise ratio (SNR) from \textit{Kepler} as listed by the NASA Exoplanet Archive. We use the Pre-search Data Conditioning (PDCSAP) light curves \citep{2012PASP..124.1000S,2012PASP..124..985S}.

The period ranges in this set are 0.7--385\,d with a mean (median) of 9.6\,d (29.4\,d). The transit durations cover the range 0.034--0.67\,d (mean 0.18\,d, median 0.15\,d). SNRs range from 6--15 (mean 12.6, median 13).

\subsection{Injected sample of real planets in young stars from \textit{TESS}}
Young star systems ($\lesssim$\,Gyr) are important probes to study the primary processes of planetary evolution, in particular if they are host to transiting planets. They serve as benchmark systems to get insights into planet formation, e.g. the migration of hot Jupiters \citep{2018arXiv180606601H}, the evolution of planetary atmospheres \citep{Madhusudhan2016}, and other physical properties such as mass-radius relationships.

The \textit{K2} and the \textit{TESS} missions monitored several young open clusters such as 
the Pleiades \citep[\textit{K2} C4, age 110\,Myr,][]{2017A&A...599A..32V},
the Hyades \citep[\textit{K2} C4 \& C13, 800\,Myr,][]{2015ApJ...807...58B},
and Praesepe \citep[\textit{K2} C5 \& C16, 800\,Myr,][]{2015ApJ...807...24B}). 
In addition, \textit{K2} observed the
Taurus-Auriga star-forming region \citep[\textit{K2} C4 \& C13, 1--5\,Myr,][]{2017ApJ...838..150K} and the 
Upper Scorpius subgroup of Sco-Cen \citep[\textit{K2} C2 \& C15, 10\,Myr,][]{2012AJ....143...72M}.

Fewer than 15 transiting planets have been discovered in young clusters and associations, although thousands of light curves have been searched \citep[e.g.,][]{2016ApJ...818...46M,2016AJ....152...61M,2016Natur.534..658D,2016AJ....152..223O,2016MNRAS.463.1780L,2017AJ....153...64M,2017AJ....153..177P,2019arXiv190209670D}.

Only a few planets are known transiting young active stars, not enough for our experiment. This low number may be in part due to the fact that light curves of young stars feature periodic variability of typically 1-5\,\% in amplitude, usually phased at the stellar rotation period. Often, significant aperiodic variability is present in addition. The (mostly rotational) variability is often temporally complex, with evolution over the course of a $\sim70$\,d \textit{K2} observing campaign \citep{2017AJ....154..224R}.

We extracted light curves for 316 high-probability members of known young moving groups and clusters \citep[]{2018ApJ...856...23G} from the \textit{TESS} Full-Frame Images using the \texttt{eleanor} pipeline \citep[]{feinstein19}. Using different methods, we performed a transit search followed by by-eye vetting for each target. No new planet candidates were detected. Therefore, we treat these light curves as free of significant transits and use them for an injection and retrieval experiment.

We injected planetary transits based on a \citet{2002ApJ...580L.171M} model using the \texttt{batman} package \citep{2015PASP..127.1161K}, with $P$ drawn randomly from the range [1, 15]\,d, a transit impact parameter of zero, an orbital eccentricity of zero, and $R_P/R_{*}$ set so that in each case the planetary radius corresponds to a $0.5\,R_{\text{Jup}}$ planet. Stellar parameters were pulled from the \textit{TESS} input catalog \citep{2018AJ....156..102S}.

To limit the computational effort, only a subset of the most promising detrending methods was tested. We chose to compare the performance of a sliding median and biweight, the Lowess, and a robust (Huber) spline. In addition, we added a new method owing to the strong periodic variability seen in many young stars. First, we ran a Lomb-Scargle periodogram. Then, with the most significant period, we informed a Gaussian Process with a quasi-periodic kernel. A Matern kernel was added in addition to capture the remaining non-periodic variation.

\begin{figure}
\centering
\includegraphics[width=\linewidth]{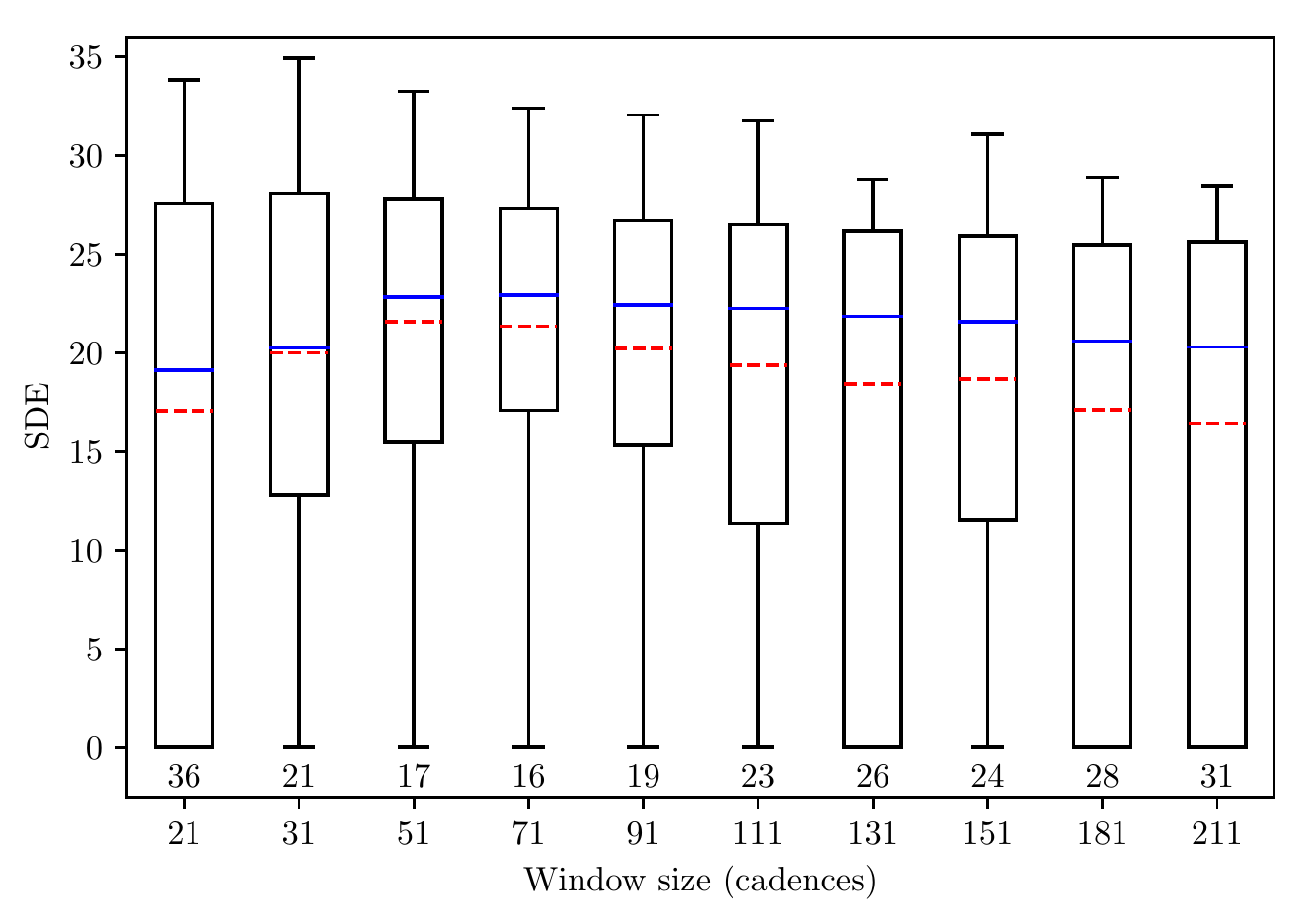}
\caption{Boxplot for different window sizes of the Savitzky-Golay filter for a {\tt TLS}-search in the \textit{Kepler} \textit{K2} sample. With polyorder $p=2$, a window size of $w=71$ cadences (horizontal axis) yields the lowest number of non-detections (16, bottom row), the highest median SDE (dashed red line) and the second-highest mean SDE (blue line). Higher polyorders yield strictly worse results (not shown). Thus, $p=2$, $w=71$ is the optimal parametrization of the Savitzky-Golay filter for a {\tt TLS}-based transit search in \textit{K2} data. Boxes cover the lower to upper quartiles, whiskers show the 10 and 90 percentiles.}
\label{fig:savgol}
\end{figure}

\subsection{Optimal parameter determination for each algorithm}
To allow for a fair comparison of the individual algorithm, the parameters of each method were optimized individually for the sample at hand. The optimization goal was to maximize the number of detections in the sample (as defined in Section~\ref{sub:performance_measurement}). For algorithms with multiple degrees of freedom, all combinations were explored extensively. As an example, we show the result for the Savitzky-Golay method in Figure~\ref{fig:savgol}. This method has two parameters: The width of the sliding window $w$ (in cadences), and the degree of the polynomial ($p$). For $p=2$, a global maximum is found near $w=71$ (as the number of missed detections, 16, is minimal), and a subsequent finer grid around this value (not shown) confirms this location. For $p=3, 4, 5, ...10$ (and many trial windows), the number of misses increases and thus $p=2$, $w=71$ is the best setting for a Savitzky-Golay filter for this particular sample.

The window size is a trade-off between local removals of trends, versus the protection of transit-like signals. With the transit duration being mainly a function of planetary period, the window size must be increased when searching for longer transits. Thus, the optimal window size is determined by the period distributions in our experiments (see also section~\ref{sec:tests}).

Our visualizations make use of the boxplot, where the box extends from the lower to upper quartile value. The median and mean values are indicated with a blue line and a red dashed line. The whiskers extend to the [10, 90] percentiles. Between the 10 percentile and the zero SDE value, we print the number of missed detections (if any).

\section{Results}
\label{sec:results}
\label{sub:performance_measurement}
We now compare the results of each algorithm using their optimal settings per dataset, unless noted otherwise.

\begin{figure*}
\includegraphics[width=\linewidth]{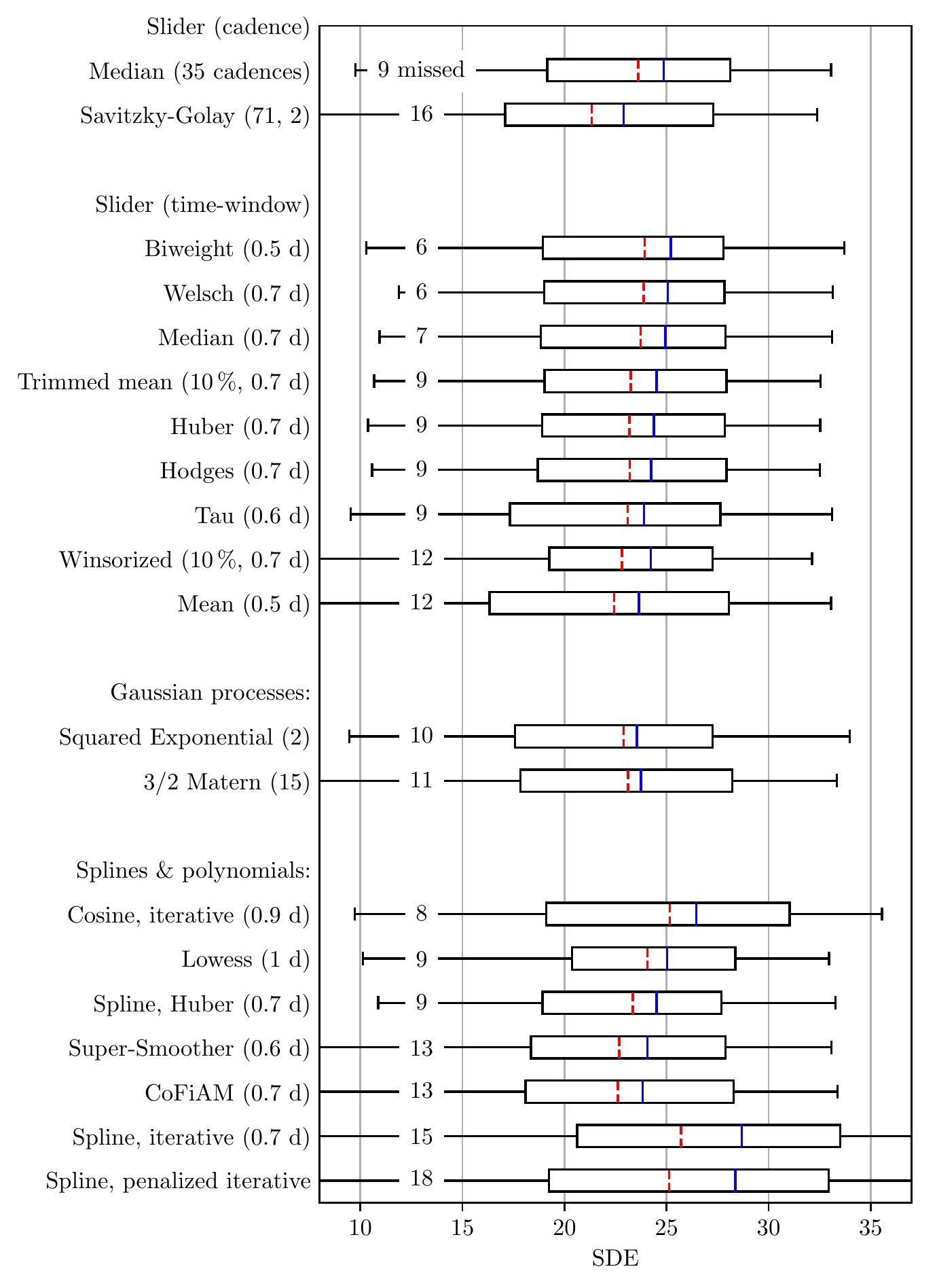}
\caption{\textit{K2} robustness experiment. Boxplot of all methods, using optimal parameters for each. Boxes cover the lower to upper quartiles, whiskers show the 10 and 90 percentiles. Dashed red lines indicate the median SDE, blue lines the mean SDE. The number left of the box is the number of missed detections (of 100). In the first group of cadence-based sliding windows, the median {\tt scipy.medfilt} with a width of 35 cadences recovers 91 of 100 planets, but misses 9 (text on the upper left). The winner of this experiment is a time-windowed slider with a robust estimator of type Tukey's biweight (or Welsch), missing only 6 planets. }
\label{fig:comparison}
\end{figure*}

\subsection{Robustness sample of real planets in \textit{K2}}
\label{sub:k2}
The most severe data quality issue in our \textit{K2} experiment were low outliers caused by the rolling of the spacecraft due to the loss of a reaction wheel. The resulting instrumental flux drops are only partially corrected in the {\tt EVEREST} pipeline. While high outliers can be removed at the small cost of skewing the distribution of data points around the normalized (or local) flux, low points cannot simply be removed -- they are often the signal we seek. If these low outliers are kept as is, $\sim20$\,\% of our simulated transiting planet sample is undetectable. We have determined an optimal threshold to clip low outliers. By testing each algorithm (using individual optimal parameters) against different values of $\sigma$ with an outlier clipping below various multiples of the standard deviation, a clipping of outliers that were $10\,\sigma$ below the mean after detrending turned out to be optimal. In this case, only $\geq6$ planet were missed in our injection-retrieval experiment, depending on the detrending. All planets can be recovered using the ``biweight'' detrending in two runs, one without clipping, and one with $10\,\sigma$ clipping. This is not true for all other detrending methods, which still miss some planets in this two-pass scenario. Most important, the ranking of methods does not change for different choices of the clipping threshold in multiples of $\sigma$.

In the reference 1-pass scenario, where the $10\,\sigma$ clipping has shown to be optimal (Figure~\ref{fig:comparison}), the ``biweight'' and ``Welsch'' location estimators are the winners with 6 misses (of 100). The time-windowed median slider is only marginally worse (7 misses). A direct comparison of the time-windowed median to a cadence-based median of the same (average) length shows that the cadence-based version misses two planets more (9 instead of 7), and its mean and median SDE values are slightly worse ($\Delta{\rm SDE}\sim0.1$): The time-based slider is superior to the (typically used) cadence-based version. Other time-based sliders are worse in performance, namely the trimmed mean, the Hodges, the Huber, and the winsorized mean. The most suboptimal choice in this group is the sliding mean, which is strongly affected by outliers (e.g., in-transit points). Yet, while missing 12 planets, a sliding mean is still superior to the Savitzky-Golay, which misses 16. 

The comparative performance of Gaussian processes is worse than all robust methods. Missing 10-11 planets (depending on the kernel used) yields a mediocre performance. Interestingly, the result does not improve for the robust (iterative) GPs. It did, however, change slightly the specific planets to be found, by changing the effective low-point outlier clipping. 

Finally, splines and polynomials perform inferior to time-windowed methods. Interestingly, some methods achieve a higher mean SDE compared to the ``biweight''. These methods perform well for highly significant transits and allow for a detection with even higher confidence. However, they miss more of the difficult planets, something we consider the more important metric. 

We have repeated the experiment using the {\tt K2SFF} reduction from \citet{2014PASP..126..948V}, which is noisier. Out of the 100 light curves with injected transits, the highest number of recoveries is 78, again achieved by the ``biweight`` method. The other methods perform only marginally worse, e.g., ``lowess`` recovers 76. This indicates that the leading order difficulty in transit recovery from {\tt K2SFF} data is noise, and not the detrending method -- at least for our sample. Using {\tt EVEREST}, however, the detrending method is very important.

\begin{figure}
\centering
\includegraphics[width=\linewidth]{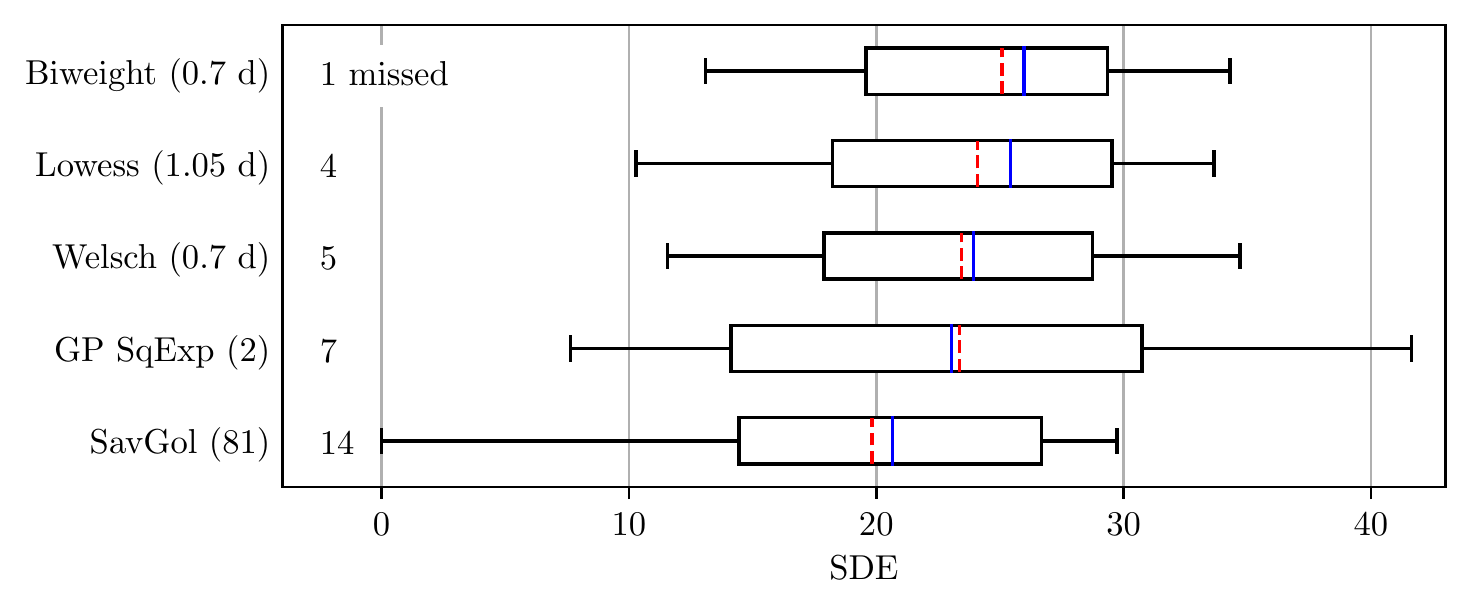}
\caption{\textit{Kepler} experiment showing boxplots and missed recoveries (e.g., 1 for the biweight) for a subset of all methods, due to computational expense. Again, the biweight slider is the most suitable method, while the performance of of GPs and the Savitzky-Golay filter are substantially worse. Boxes cover the lower to upper quartiles, whiskers show the 10 and 90 percentiles. Dashed red lines indicate the median SDE, blue lines the mean SDE.}
\label{fig:boxplot_k1}
\end{figure}

\subsection{Performance sample of shallowest planets in \textit{Kepler}}
Test results from this subset are very similar to the \textit{K2} sample. In the group of sliders, we find that the biweight is slightly better than the Welsch and the median, and better by a large margin than the other estimators. The Savitzky-Golay shows a very weak performance, from which we conclude that this issue is not caused by \textit{K2} systematics. Instead, this method should be avoided for any light curve preparation before a transit search, as it removes a relevant fraction of transits (10--20\,\%). In the group of Gaussian processes, we tested only a squared exponential kernel, again with a mediocre performance. In the subgroup of splines and polynomials, we find again Lowess with an acceptable performance, although still inferior to the best robust slider. Figure~\ref{fig:boxplot_k1} shows a performance summary of the relevant methods.

Overall, it appears that the relative sensitivity (i.e. recovery rate) and robustness (i.e. SDE) of the detrending methods do not depend significantly on the data quality. Initially, we had expected that some methods might exhibit better performance than others in the presence or absence of strong outliers. This is not the case in our experiment. We were not able to extract a change of preference of the detrending methods depending on whether we explored \textit{K2} data with strong systematics or \textit{Kepler} data with the most shallow transits.

\begin{figure}
\centering
\includegraphics[width=\linewidth]{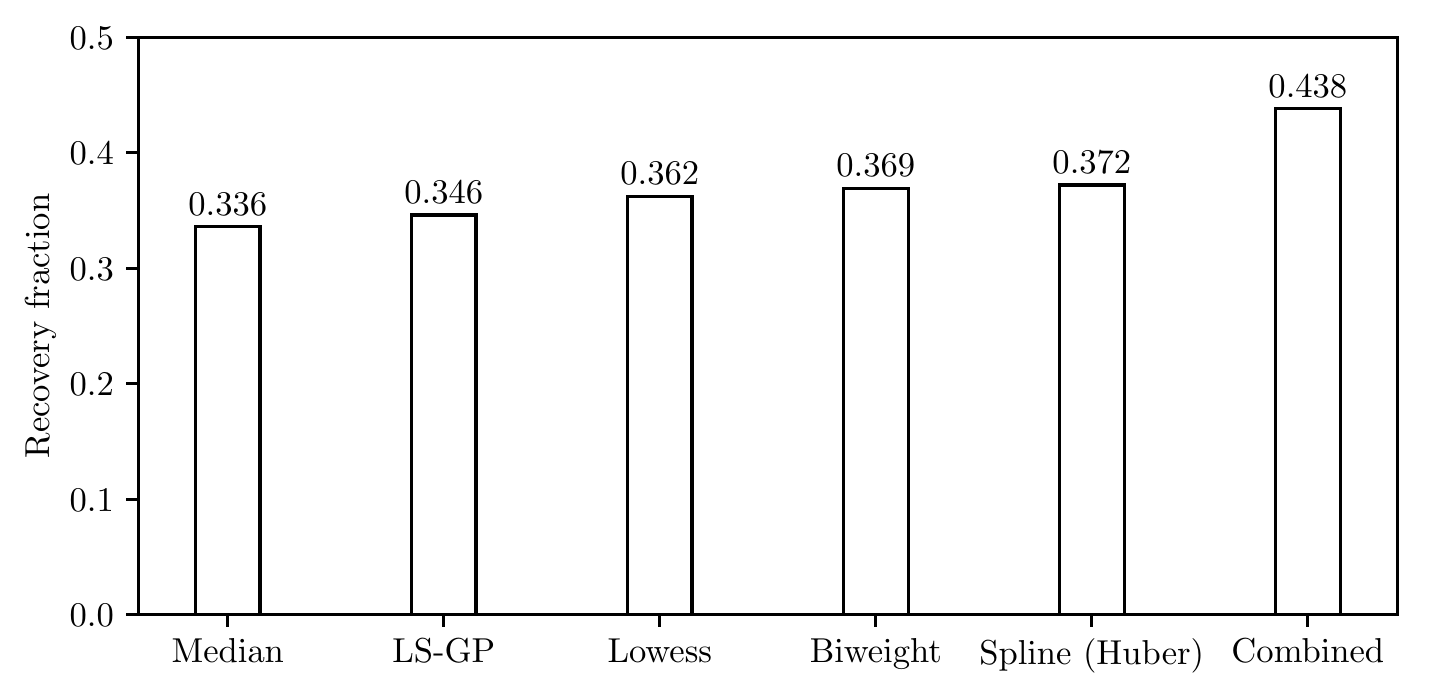}
\caption{Injection-retrieval experiment using 316 young stars from \textit{TESS}. In each light curve, a $0.5\,R_{Jup}$ central-transit planet was injected with periods in the range 1--15\,d. As stellar variability was strong and often periodic, we added a new method: a periodic Gaussian Process informed by a period pre-search (based on a Lomb-Scargle periodogram), plus a Matern kernel to capture the remaining non-periodic variation. This method was optimized by varying the kernel size. As a global maximum, it recovered 34.6\,\% of the transits. Due to computational expense, only a few methods were re-tested here. Interestingly, the robust Huber spline (with $w=0.3$\,d) slightly outperformed the sliding biweight ($w=0.25\,$d). Combining all methods increases the total yield to 43.8\,\%.}
\label{fig:young_bars}
\end{figure}

\subsection{Injected sample of real planets in young stars from \textit{TESS}}
The most surprising result from our experiment with injected transits into \textit{TESS} light curves of young stars is the low recovery rate of all methods. It ranges from 33.6\,\% for the sliding median to 37.2\,\% for the robust spline (Figure~\ref{fig:young_bars}). In less noisy data, the recovery rate of such planets with $0.5\,R_{Jup}$ is close to 100\,\%.

The second surprise is that all methods perform very similarly and that their parameter settings have little influence on the recovery rates. At first glance, this suggests that the detrending method is less important, and other factors dominate. At second glance, it is interesting that the overlap of actual planets recovered with each method is large, but not complete. Combining all 5 methods pushes the recovery rate to 43.8\,\%, at the cost of $5\times$ as many vetting sheets to be examined by humans (or a suitable robo-vetter). A typical failure mode is shown in Figure~\ref{fig:young} and discussed below (Section~\ref{sub:win_size}).

Finally, it can be noted that the sliding biweight is not the answer to all detrending questions. It performed best in the \textit{Kepler} and \textit{K2} experiment, but was beaten by the robust spline in young \textit{TESS} stars. Although the difference was marginal, it reiterates the point that the best method for a given task can (and often should) be determined quantitatively.

\subsection{Best edge treatment}
We tested the influence of the possible edge treatments (see Section~\ref{sub:edge}) using a sliding median as implemented in {\tt scipy.medfilt}, and a time-windowed biweight. The worst results (as measured in the fraction of planets recovered) were found when {\it wrapping} the data from the beginning to the end of the time series. This should work well if the data are periodic and the period is equal to the length of the time series, which is usually not the case. Padding with constant values, and mirroring, produce very similar, usually acceptable, results. Slightly better results are achieved when the sliding window shrinks near edges. The best results, however only by a slight margin of $\Delta{\rm SDE}\sim0.1$, were achieved when {\it discarding} data (half a window size) near edges.

Using more data improves the SDE if these data are generally acceptable. With difficult data, it is preferable to remove them instead. Discarding data near edges can be avoided by verifying by-eye that the trend does diverge near the borders. Usually, this is the preferred method when presenting results in publications, and Figures should include all usable data. For automatic search programs, where this decision must be made automatically, it is (slightly) preferable to discard these few points (usually of order one percent of the data).

\subsection{Best window shape for sliders}
Throughout the paper, we have used a boxcar (rectangular) window for the time-window based methods, except for the LOWESS, which uses a tri-cube. This requires a justification. Other window options, such as a Gaussian or an exponentially-weighted moving estimator, concentrate more weight towards the central moment. With correlated noise, this should improve the efficiency of the location estimate, as correlated noise decays over some finite time. In practice, however, too much of a concentration towards the center is futile. During long-duration transits, the central moment falls into the transit dip, so that the trend fits out the actual transit. This can be compensated by making the window longer, and/or stronger emphasis on the robustness. Longer windows, however, leave substantial residue after detrending when the stellar noise periodicity is short. We have tested a subset of the robust sliders (the biweight, the mean, and the median) with Gaussian and Epanechnikov windows. In all cases, the recovery results were worse. Thus, we recommend to use the rectangular window or, as the case may be, the tri-cube weighting as in the LOWESS method.

\begin{figure}
\centering
\includegraphics[width=0.7\linewidth]{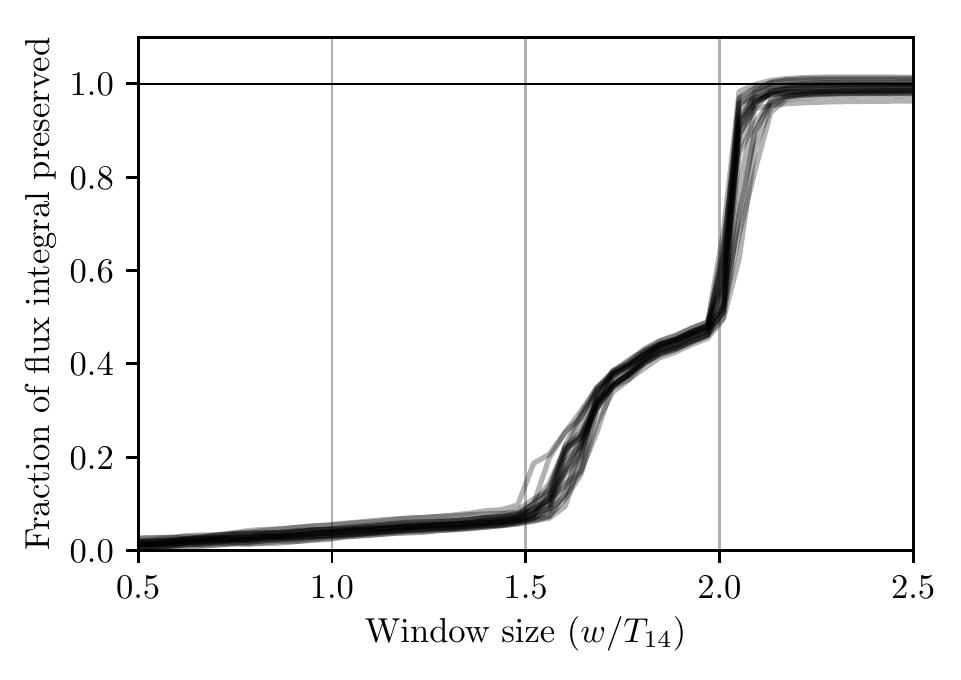}
\caption{Fraction of the flux integral preserved, as a function of the window size (in units of $T_{14}$) for a sliding biweight. Data from an injection-retrieval experiment of Mandel-Agol transit models into Gaussian noise where the transit depth is $10\,\times$ the standard deviation per data point. For $w/T_{14} \gtrsim 2.2$, most ($\gtrsim 98\,$\%) of the flux integral is preserved.}
\label{fig:mc}
\end{figure}

\begin{figure*}
\centering
\includegraphics[width=.49\linewidth]{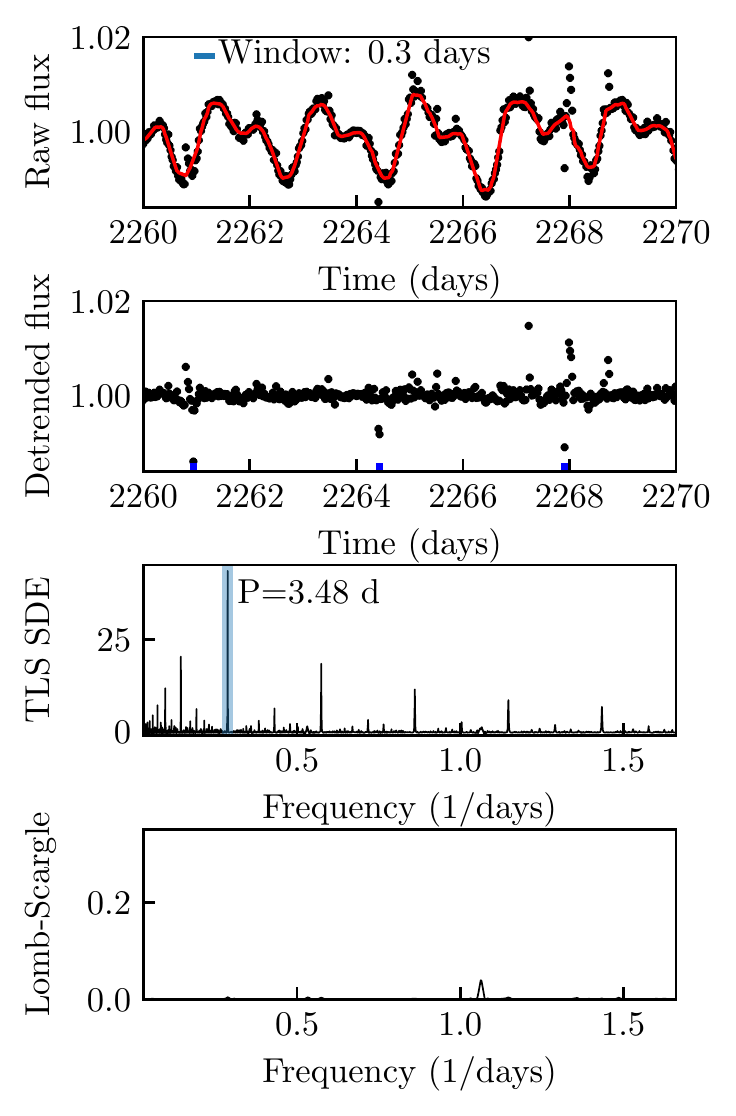}
\includegraphics[width=.49\linewidth]{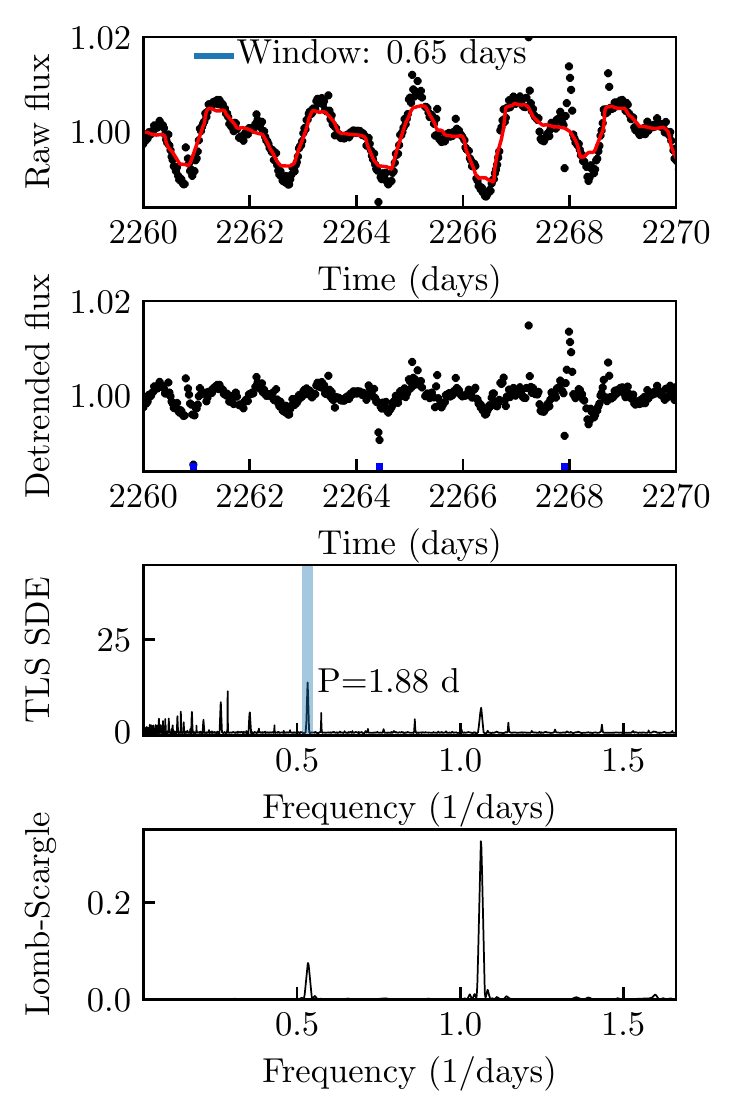}
\caption{Detrending of K2-25\,b \citep{2016AJ....152...61M} with the biweight detrending filter and two different window sizes: short (0.3\,d, left panels), which allows for a detection of the planet ($P=3.4843\,$d), long (0.65\,d, bottom panels). From top to bottom: Raw flux, detrended flux (both only shown for a segment of the total time series), {\tt TLS} signal detection efficiency, Lomb-Scargle periodogram. The longer window leaves stellar noise at a level so that the automatic transit search detects the stellar variation instead of the transit as the most significant signal. The shorter window, however, would remove planets where $T_{\rm 14} \gtrsim 2w$, which is $P \gtrsim 12\,$d for this $M=0.3\,M_{\odot}$, $R=0.3\,R_{\odot}$ star and a $R_{\rm Jup}$ planet.}
\label{fig:young}
\end{figure*}

\subsection{Optimal window length}
\label{sub:win_size}
The window length (kernel size, knot distance...) should be as short as possible to maximize the removal of stellar variability (Figure~\ref{fig:young}), but as long as necessary to preserve the transits. With another experiment, we now determine the optimal window size for the time-windowed biweight slider. 

In this separate experiment, we re-use the \textit{Kepler} sample of the 100 lowest SNR planets. Now, the window size was set {\it for each planet individually} as a multiple of its published transit duration $T_{14}$, as queried from NASA Exoplanet Archive. As shown in Figure~\ref{fig:bi_k1}, the highest minimum (and mean) SDE is achieved for a window of size $3T_{14}$. This choice recovers all planets. We have also checked the limiting case of white noise using synthetic injections and retrievals into Gaussian noise. Then, the window can be slightly shorter ($w/T_{14}\gtrsim 2.2$) to preserve almost all ($\gtrsim 98\,\%$) of the flux integral (Figure~\ref{fig:mc}).

If we instead use a fixed window for all stars, we may reasonably decide to choose it so that the planet with the longest transit duration is still preserved, which is $T_{14} \sim 0.67\,$d for Kepler-1638\,b with $P=259\,$d. Selecting $w=2.5T_{14} \sim 1.65\,$d, we run another search with this window for all stars in the sample. The recovery fraction is now only 87\,\%, losing 13 planets. Most of these have short periods (with corresponding short transit durations) and strong stellar variation, which is not sufficiently removed with a 1.65\,d window. This shows that an adaptive window (per star, per period, per transit duration) is superior to a single fixed size. But there is a problem: For a blind search, the period is unknown, and thus $T_{14}$ is unknown. However, we can estimate the transit duration it as a function of period given stellar mass and radius. As a sensible limit, we can use a large planet ($2R_{\rm Jup}$) and a circular orbit, and find \citep[see Eq.~(10) in][]{2019A&A...623A..39H}

\begin{equation} \label{eq:T14max_2}\nonumber
T_{14,{\rm max}} = (R_{\rm s}+R_{\rm p}) \left( \frac{4P}{\pi G M_{\rm s}} \right)^{1/3}.
\end{equation}

With $w/T_{14}=2.2$ we have $w=1.54\,$days for the longest periods, but $w=0.2\,$days for a short period ($P=1\,$d) (the longest transit durations from the \textit{Kepler} mission are actually $T_{14}\sim1\,$d \citep[e.g., Kepler-849,][]{2016ApJ...822...86M}). Simply choosing a unique long window length for all searches would result in sub-optimal detrending, leading to reduced sensitivity: For the shorter-period planets, the sliding window can be $10\,\times$ shorter. In the future, improved transit search algorithms should include a custom detrending for each period. The resulting gain in the computational expense could be mitigated by defining clusters of ranges of periods.

\section{Discussion}

\subsection{Which method to choose?}
The subject of this paper is the removal of stellar trends with simultaneous preservation of transit signals that shall be detectable with {\tt TLS}. Our results may apply to other problems as well, where the signal occupies small segments of a time series of measurements. Other problems such as asteroseismology, however, are so different in nature or phenomenology from the transit problem considered in this work that the methods explained here may be inapplicable. For example, when filtering light curves with the aim to find phase curve variations, the window size must be much longer than for transit searches (but short enough to remove stellar rotation). Again, the right detrender can be chosen through a comparison of several methods based on simulated or partly simulated data as done in this paper.

\begin{figure*}
\centering
\includegraphics[width=\linewidth]{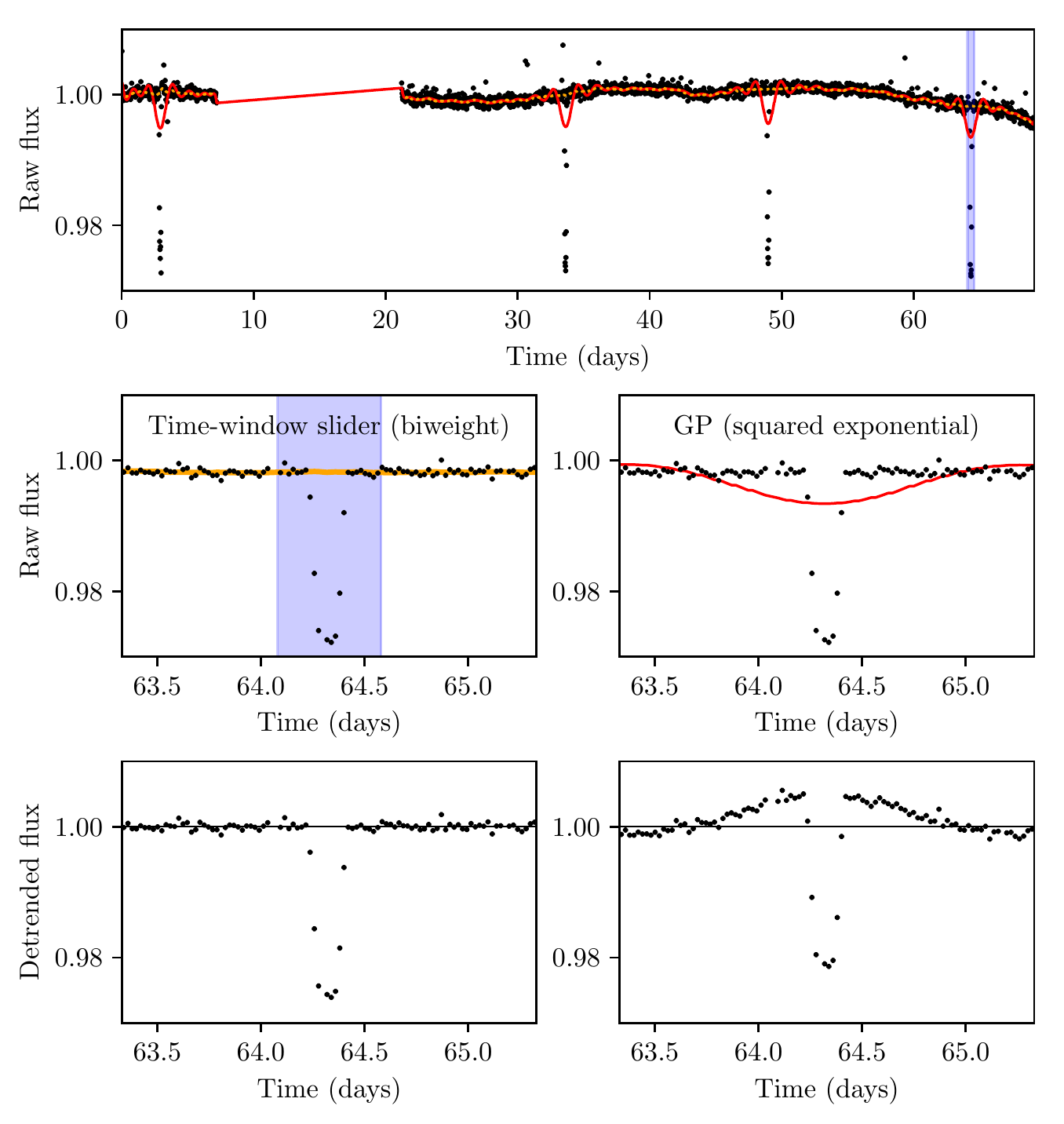}
\caption{Detrending for EPIC 228707509 \citep{2018AJ....156...78L} using the time-windowed biweight (top and left, orange lines) with its optimal window size of 0.5\,d (blue shade), and a Gaussian Process with a squared exponential kernel of optimal size (top and right, red lines). \textit{K2} data are shown with black points, the detrending functions are illustrated as red lines. The detrending function determined with GP exhibits a substantial long-trend dip caused by the deep transit, which creates extra noise in the detrended light curve and reduces sensitivity in a transit search. An iterative approach reduces the problem in this specific example, but does not eliminate it. Overfitting can be avoided in cases where the transit ephemeris is known, by masking the in-transit points.}
\label{fig:gp_bi}
\end{figure*}

\subsection{Issues with Gaussian Processes}
GPs have performed better than a sliding mean, but worse than robust estimators including a simple sliding median in this experiment. Why is that the case?

The fact that the performance of GP as a detrender is similar to that of the sliding mean gives a first clue to the problem. Gaussian processes are built, as their name suggests, on the assumption of data which has a Gaussian distribution. This is obviously not the case in the presence of substantial short-term astrophysical activity such as transits. The GP has no way of ``knowing'' that the in-transit dip is not part of the trend in the light curve that it is supposed to fit. Based on least-squares calculations, the problem becomes more severe for stronger signals (see Figure~\ref{fig:gp_bi} for a visualization). There are two choices to tackle this issue.

The first is the classical clip-and-repeat method, similar to a multi-pass trimmed mean, where outliers are clipped and the fit is repeated until convergence. Of course, this clipping is applied only temporarily to estimate the {\tt trend}: When dividing the raw data by the trend, the in-transit points are still present. Compared to other robust (single-pass) methods, the issue is that not only the in-transit points are clipped (the desired behaviour), but also some of the points before ingress and after egress (middle panel in Figure~\ref{fig:gp_bi}). This reduction of (otherwise good and useful) out-of-transit data points decreases the fit quality. It can be visually seen in this example, but is usually invisible in practice, where individual transits are drowned by noise (but still present). This is why iterative GPs (and splines) are inferior to robust methods for the purpose of transit searches.

Other methods to make GPs more robust are currently being studied. Our literature review revealed a range of methods such as the mixture of two Gaussians \citep{Tresp01mixturesof,NIPS2001_2055}, kernel ridge regressions, and outlier identification \citep{DBLP:journals/jmlr/KussR05}. Since the essential structure of the GP remains unchanged, in all these cases the noise is assumed (near) normal, and none may be appropriate when errors appear with their own structure. We are not aware of any work that explicitly targets the problem of clustered outliers.

\begin{figure}
\centering
\includegraphics[width=0.7\linewidth]{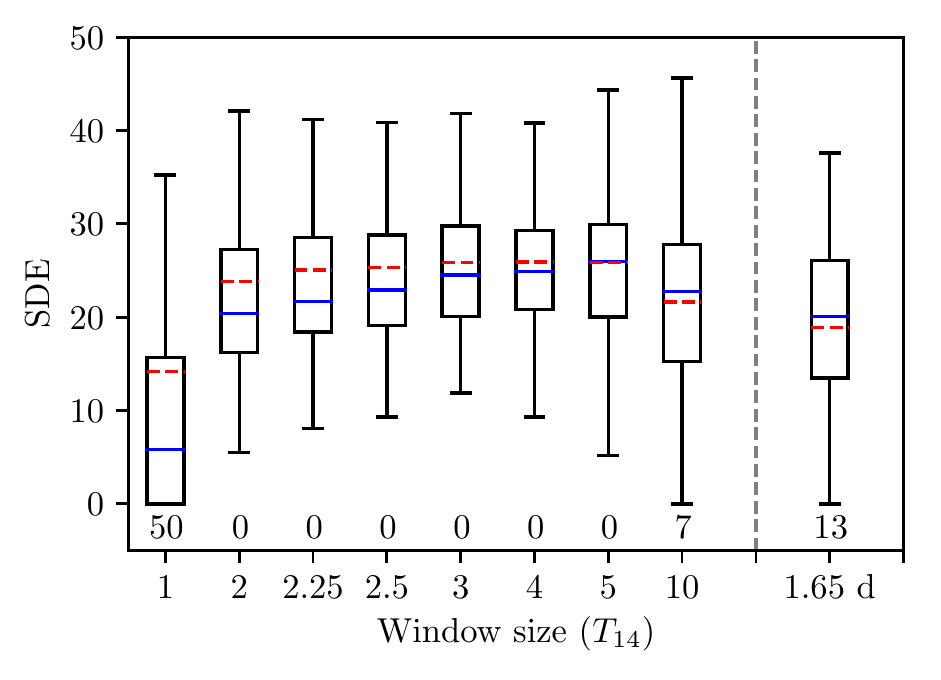}
\caption{\textit{Kepler} sample of the 100 lowest SNR planets, searched with {\tt TLS} using the sliding biweight. In this separate experiment, the window size was adjusted {\it for each planet individually} as a multiple of its known transit duration $T_{14}$. The highest minimum (and mean) SDE is achieved for a window of size $3T_{14}$. The last boxplot shows an alternative experiment, where the window is fixed to 1.65\,d for all planets, because 1.65\,d is $2.5\times$ the transit duration of the planet with the longest transit duration in the sample (0.67\,d, Kepler-1638b, $P=259\,$d). Clearly, an adaptive window (per star) is superior to a fixed size. Boxes cover the lower to upper quartiles, whiskers show the 10 and 90 percentiles. Dashed red lines indicate the median SDE, blue lines the mean SDE. Numbers at the bottom show the percentage of missed planet detections.}
\label{fig:bi_k1}
\end{figure}

\begin{figure*}
\centering
\includegraphics[width=0.48\linewidth]{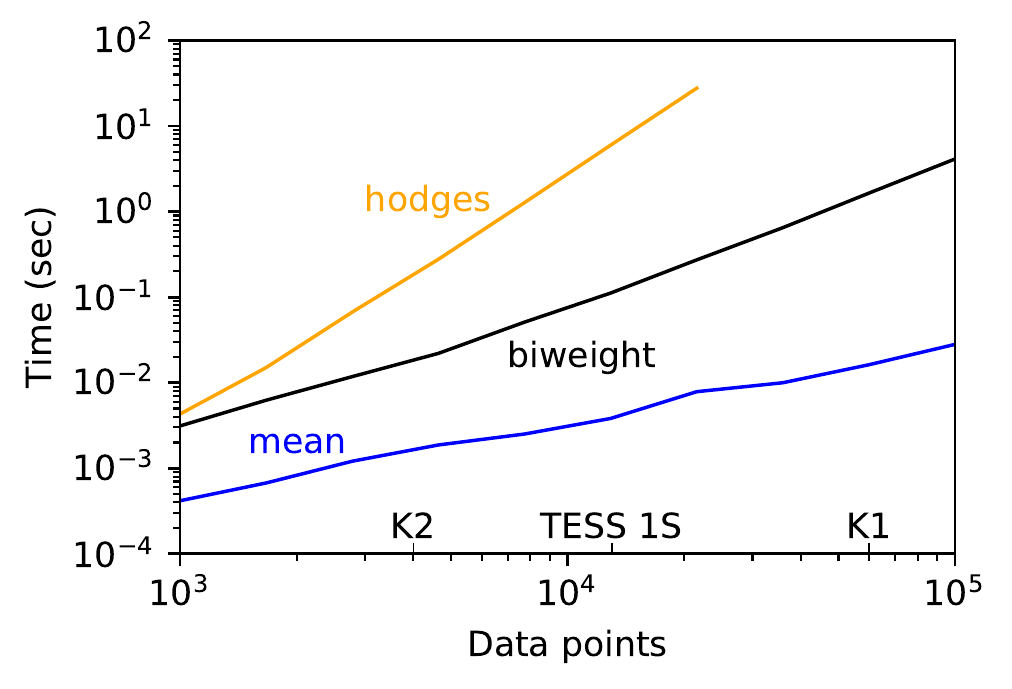}
\includegraphics[width=0.48\linewidth]{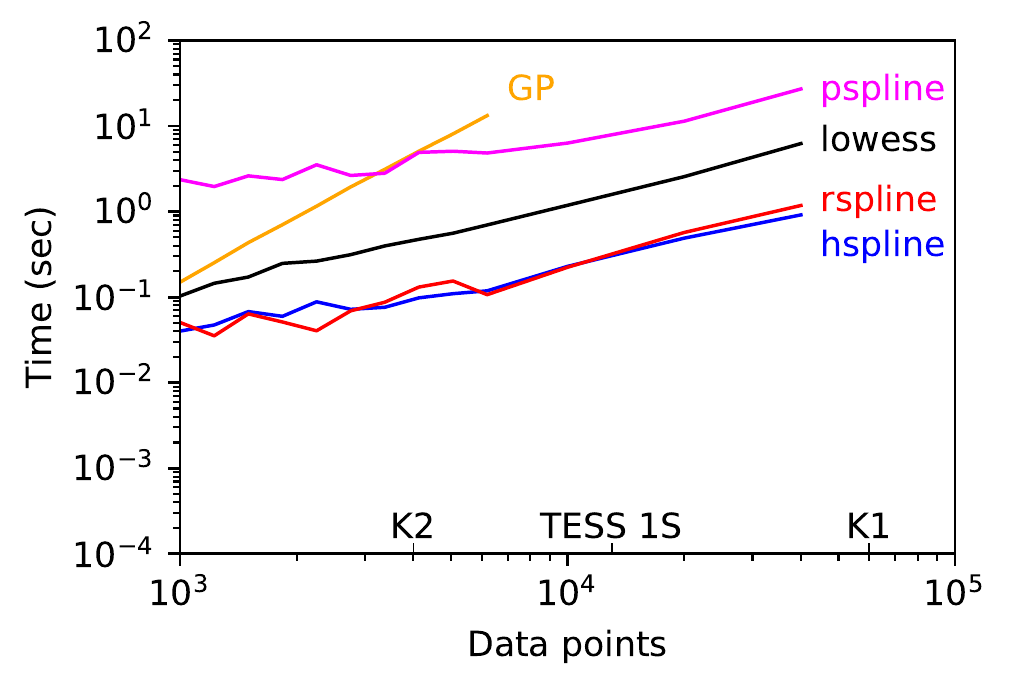}
\caption{Run-time comparison of various detrending algorithms. Left: Sliders. Right: Lowess, GP and splines (pspline: penalized, rspline: robust iterative, hspline: Huber estimator).}
\label{fig:speed}
\end{figure*}

\subsection{Computational expense}
\label{speed}
The computational expense required for our search of hundreds of transits repeatedly throughout the full parameter space of various detrending methods was significant. While most of the computing power was devoted to the actual transit search, it is crucial to assure that the detrending itself is as fast as possible, because it is performed for millions of light curves in current and future surveys.

We have measured the computational speed of the implementations used in this work (Figure~\ref{fig:speed}). The time-windowed sliding mean is the fastest, despite its naive implementation where it is re-calculated at each position (similarly to the median, trimmed mean, and winsorized mean). The biweight (and similarly Andrew's sine wave and the Welsch estimator) are slower. A robust spline has higher initiation cost, but a better performance towards large data. The slider with the Hodges estimator and the GPs are of acceptable speed ($\lesssim 10\,$s) for data with $\leq 10^4$ points, i.e. \textit{TESS} 1-sector SC and \textit{K2} LC. The other methods here are fine for data with $\leq 10^5$ points, which includes the complete \textit{Kepler} LC set.

Some methods like the iterative biweight benefit from a run-time compilation which transforms the pure Python code into a lower-level machine language before execution. We chose to implement many of the time-critical parts of \wotan \, with the specialized {\tt numba} package \citep{Lam2015}. This procedure is called ``just-in-time'' compilation and brings a thirtyfold gain in performance.

Generally, many methods can be implemented in a faster way if required. It is straightforward to see that a sliding mean filter does not require a re-calculation of the points inside the window at each move. With one value entering the window, and one value being dropped, it is trivial to re-calculate the mean efficiently. For the median, however, naive (but common) implementations {\it do recalculate} the in-window points. It was shown by \citet{Hardle1995} that this is not necessary. Instead, the values in the window need to be sorted only once. Afterwards, each new (dropped) value is inserted (removed) at the known spot. This operation can be done in $\mathcal{O}(\log_2{n})$ using the bisection algorithm \citep[][6.2.1]{knuth1998art}. Open-source implementations are available\footnote{e.g., \url{https://github.com/suomela/median-filter}} \citep{2014arXiv1406.1717S}. This method is commonly known as the ``Turlach'', after the implementation by Berwin A. Turlach\footnote{\url{https://svn.r-project.org/R/trunk/src/library/stats/src/Trunmed.c}} (1995, unpublished). Interestingly, this method is re-invented regularly \citep[e.g.,][]{2014MNRAS.445.2698H}. It has also been shown that a robust (truncated) sliding mean (robust only in $x$) is possible \citep{2016arXiv160108003J}, but it is unclear if it is possible to extend it for $y$ and/or other robust estimators.

In summary, most algorithms implemented in \wotan\,\,scale as $\mathcal{O}(n)$ for small window sizes, and $\mathcal{O}(n^2)$ if the window is a significant fraction of the data. The only exceptions are the Hodges slider (which is in $\mathcal{O}(n^3)$) and GPs ($\mathcal{O}(n^3)$ or $\mathcal{O}(n\,\log^2\,n)$, see \citet{2015ITPAM..38..252A}, depending on the implementation).

\subsection{Masking known transits}
Our work is focused on exoplanet discovery, and during such a search it is unknown which data points are in-transit (if any). Thus, the trend removal filter should cause minimal distortions to the transit shape, while removing as much stellar and instrumental variation as possible. All filters tested here achieve this task only imperfectly. Some filters, such as the biweight, are more resistant to fitting out transits than others (such as the Savitzky-Golay). Still, any filter will, on average, cause \textit{some} (usually unwanted) distortion on the transit shape.

Such distortions can be avoided (or at least maximally minimized) by masking in-transit points when known. In a post-detection analysis of the planet parameters, it is best practice to mask the in-transit points \citep[see e.g. Figure 4 in][]{2018AJ....156...99L}. \wotan{ }offers a \texttt{transit\_mask} feature for that use case.

\subsection{The way into the future}
While we have given a comprehensive overview of the most widely used detrending methods for stellar light curves, there are certainly further, less well known filters that could be tested. For example, \citet{2017AJ....154..224R} fit a polynomial plus a periodic notch-filter, which is a box-shaped region of data points that are masked from the trend. By iterating over the notch width, period, and phase, planetary in-transit points can be blanked so that they do not affect the fit. A Bayesian model selection is used to determine the filter. This is a half-way towards a simultaneous search and detrending, with the search being a box-shaped filter \citep[BLS,][]{2002A&A...391..369K}. Such an approach has been tested by \citet{2015ApJ...806..215F}. A disadvantage of this method is that the computational expense is very high, but with Moore's Law at work \citep{moore1965cramming}, it may be a beneficial approach for young, very active stars.

To further improve splines, current research is making progress towards L1 smoothers which are tailored for skewed distributions \citep{rytgaardstatistical}.

By having optimized planet-finding using {\it known} planets, we may have biased the optimization towards their parameter space. For example, we have estimated transit duration assuming circular orbits, whereas eccentric planets can have longer (or shorter) transit durations. Therefore, it can be justified to choose parameters outside of the ``optimal'' ranges listed here, for experimental and explorative purposes \citep{Loeb2010}. More philosophically: Adjusting a boat to perform well in the oceans we know does not necessarily make it fit for all unknown waters, whose nature we do not know. As of now, however, it is the best we can do. Finding more planets in the known parameter space is a good start.

In the future, it may be possible to adjust the method and/or their parameters on a per-light curve basis using machine learning \citep{2018AJ....156....7H,2018MNRAS.474..478P}. With such an approach, over- and underfitting could be reduced, increasing the planet yield. Alternatively, one may consider to circumvent detrending entirely and use machine-learning to search for transits.

\section{Conclusion}
We have investigated, for the first time, the relative performance of various detrending methods of stellar light curves with stellar activity and outliers on the detectability of exoplanetary transits. We devised and executed an injection-retrieval experiment of simulated transits into \textit{Kepler}, \textit{K2}, and \textit{TESS} light curves with a simulated search using the {\tt TLS} algorithm. As a consequence, our results can be used to efficiently search for transits in stellar light curves from these surveys. Determining empirically which tool works best for a given job should be the norm, and our framework can serve as an example. Our toolkit for time-series detrending, dubbed {\tt W\={o}tan}, is open source under the MIT license and is available at \url{https://github.com/hippke/wotan} with documentation and tutorials. It can serve as a one-stop solution for various smoothing tasks, and its open source nature gives the possibility to the community to contribute improvements.

Based on our experiments for transit detection,  we find that a time-windowed slider with an iterative robust location estimator based on Tukey's biweight is an optimal choice in many cases. This is usually superior to the median or trimmed methods. In the presence of a lot of stellar variation, spline-based methods work comparably well. We also determined that the optimal width of the sliding window is about three times the transit duration that is searched for.

\software{
eleanor \citep{feinstein19},
numpy \citep{numpy:2011},
numba \citep{Lam2015},
scipy \citep{scipy:2001},
scikit-learn \citep{scikit-learn},
astropy \citep{2013A&A...558A..33A,2018AJ....156..123A},
statsmodel \citep{seabold2010statsmodels},
george \citep{2015ITPAM..38..252A,2017AJ....154..220F},
celerite \citep{2017ascl.soft09008F,2018RNAAS...2a..31F},
pyGAM \citep{danielserven},
matplotlib \citep{matplotlib:2007},
iPython \citep{PER-GRA:2007},
Jupyter Notebooks \citep{Kluyver:2016aa},
TLS \citep{2019A&A...623A..39H}
}

\acknowledgments
We thank Adina D. Feinstein for assistance with TESS light curves, Kai Rodenbeck and Alex Teachey for providing code to implement the {\tt CoFiAM} algorithm, David Kipping for corrections about the details of the CoFiAM method, Andrew Vanderburg for help to with K2SFF files, and Aviv Ofir for ideas on a sum-of-sines method with iterative sigma clipping. This project was developed in part at the {\it Building Early Science with \textit{TESS}} meeting, which took place in 2019 March at the University of Chicago. RH is supported by the German space agency (Deutsches Zentrum f\"ur Luft- und Raumfahrt) under PLATO Data Center grant 50OO1501. T.J.D. gratefully acknowledges support from the Jet Propulsion Laboratory Exoplanetary Science Initiative. Part of this research was carried out at the Jet Propulsion Laboratory, California Institute of Technology, under a contract with NASA.

\bibliography{references}
\end{document}